\documentclass[letterpaper,english,reprint, aps]{revtex4-1}
\usepackage[T1]{fontenc}
\usepackage[latin9]{inputenc}
\usepackage{color}
\usepackage{float}
\usepackage{amsmath}
\usepackage{amssymb}
\usepackage{graphicx}

\makeatletter

\pdfpageheight\paperheight
\pdfpagewidth\paperwidth

\makeatother

\usepackage{babel}
\begin{document}

\preprint{APS/123-QED}

\title{Counter-propagating charge transport in the quantum Hall effect regime}

\author{Fabien Lafont$^{1,2,*},$Amir Rosenblatt$^{1,*},$ Moty Heiblum$^{1},$
Vladimir Umansky$^{1}$}

\affiliation{$^{1}$Braun Center for Submicron Research, Dept. of Condensed Matter
physics, Weizmann Institute of Science, Rehovot 76100, Israel}

\affiliation{$^{2}$Collège de France, 11 place Marcelin Berthelot, 75231 Paris
Cedex 05, France}

\address{$^{*}$These authors contributed equally to this work}

\maketitle
\textbf{\textcolor{black}{The quantum Hall effect, observed in a two-dimensional
electron gas subjected to a perpendicular magnetic field, imposes
a 1D-like chiral, downstream, transport of charge carriers along the
sample edges. Although this picture remains valid for electrons and
Laughlin\textquoteright{} s fractional quasiparticles, it no longer
holds for quasiparticles in the so-called hole-conjugate states. These
states are expected, when disorder and interactions are weak, to harbor
upstream charge modes. However, so far, charge currents were observed
to flow exclusively downstream in the quantum Hall regime. Studying
the canonical spin-polarized and spin-unpolarized $\nu=2/3$ hole-like
states in GaAs-AlGaAs heterostructures, we observed a significant
upstream charge current at short propagation distances in the spin
unpolarized state.}}

\textcolor{black}{Elementary charge excitations in the quantum Hall
effect (QHE) flow downstream along the edge of a two-dimensional electron
gas (2DEG), with the downstream chirality imposed by the magnetic
field \citep{Klitzing1980}. In the fractional regime \citep{Tsui1982}
this statement remains valid only for particle-like (Laughlin\textquoteright s)
states \citep{Laughlin1983,Beenakker1990,WEN1991}; in contrast, hole-like
states (filling factors $\nu$ such that $1/2+n<\nu<1+n$ with $n=0,1,2...$),
are expected to harbor counter-propagating (downstream and upstream)
charge excitations \citep{MacDonald1990}. In a non-interacting and
scattering-free model, a downstream $\nu=1$ charge mode was predicted
to be accompanied by an upstream $\nu=1/3$ mode, leading to a two-terminal
conductance of $4e^{2}/3h$ where $e$ and $h$ are respectively the
electron charge and the Planck constant. However, experimentally,
only downstream charge modes \citep{Ashoori1992,Sabo2017} with a
two-terminal conductance of $2e^{2}/3h$ accompanied by upstream neutral
modes\citep{Kane1994,Meir1994,Wang,Bid2010,Gurman2012,Inoue2014a,rosenblatt2017transmission}
have been found. A recent experiment \citep{Grivnin2014} measured
conductance of an unequilibrated downstream channels at narrow regions
(4~$\mu\text{m}$ wide) of the polarized $\nu=2/3$ state; the results
were consistent with the model from \citep{MacDonald1990} but no
direct measurement of the upstream current was made. Although the
majority of the studies were concentrated on the spin-polarized $\nu=2/3$
state, there has been recent interest in its spin-unpolarized counterpart
\citep{Eisenstein1990,Kronmuller1998,Kronmuller1999,Chakraborty2000,Smet2001b,Kraus,Hayakawa2012,Moore2017}
- as a potential host for para-fermions when coupled to superconducting
contacts \citep{Mong2014,Clarke2014,Wu2018}. In the Composite Fermion
(CF) picture, one can construct two kinds of states in the $\nu=2/3$:
An unpolarized state, emerging at lower magnetic fields, with two
quantum levels that have the same orbital quantum number }
\begin{figure}[H]
\centering{}\textcolor{black}{\includegraphics[width=0.75\columnwidth]{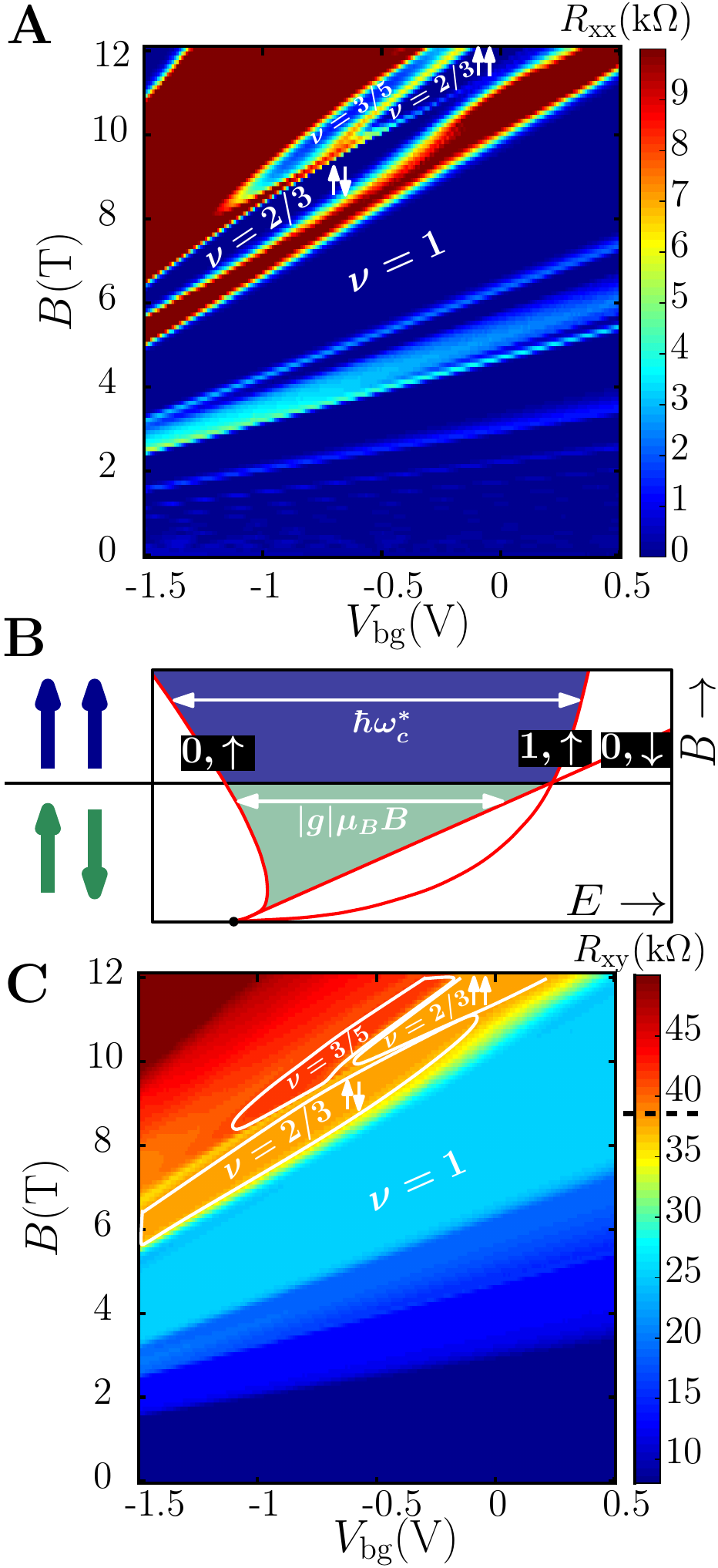}\caption{\textbf{Longitudinal and transverse magneto-resistances measured in
a }40 $\mu$\textbf{m wide Hall bar sample.} \textbf{(A)} Longitudinal
four-terminal magneto-resistance versus backgate voltage measured
using $I=$1 nA at $T=40$ mK . A clear non-dissipated state, $R_{xx}\approx0$,
is visible for the $\nu=2/3$ polarized and unpolarized states. \textbf{(B)}
Sketch of the evolution of the relevant energy scales: At low field
a gap exists between the $\left(0,\uparrow\right)$ and $\left(0,\downarrow\right)$
states corresponding to the spin unpolarized state), whereas at higher
fields, thanks to the different $B$ dependency of the Coulomb ($\propto l_{B}^{-1}\sim\sqrt{B}$,
where $l_{B}$ is the magnetic length) and Zeeman ($\propto B$) energies,
the gap exists between the $\left(0,\uparrow\right)$ and the $\left(1,\uparrow\right)$
lambda levels corresponding to the polarized state. \textbf{(C)} Four-terminal
transverse magneto-resistance as function of the backgate voltage.
The $\nu=2/3$ polarized and unpolarized quantum Hall plateaus exhibit
a resistance $R_{xy}\approx(3e^{2}/2h)^{-1}\approx38.7\,\text{k}\Omega$
(dashed line on the color bar). }
}
\end{figure}
\textcolor{black}{{} but opposite spin configurations: (0,$\uparrow$)
and (0,$\downarrow$) (Fig. 1B) \citep{Jain1989}, and a polarized
state, emerging at high magnetic fields, with two quantum levels having
the same spin but different orbitals (0,$\uparrow$) and (1,$\uparrow$)
\citep{Kukushkin1999}. The majority of previous experiments in the
unpolarized state focused on characterizing the spin domains structure
in the bulk \citep{Verdene2007,Hayakawa2012,Moore2017} or the nuclear
spin polarization occurring at high currents \citep{Smet2001,Kraus,Huels2004,Verdene2007,Hennel2016,Cho1998,Kronmuller1998,Kronmuller1999,Li2012,smet2002}.
Still, the configuration of edge channels for this state remains elusive:
on the one hand, no upstream channel is expected in the CF picture,
on the other, because the effective K-Matrix in the CF basis is the
same for both $\nu=2/3$ states, an upstream mode should occur also
in the unpolarized case \citep{Wu2012}. Here, we studied the two
flavors of the $\nu=2/3$ state along short distance (a few $\mu$m)
and found a substantial upstream charge current only in the spin-unpolarized
state. }Consequently, the two-terminal resistance deviates from the
quantized one at $\nu=2/3$. \textcolor{black}{The GaAs-AlGaAs heterostructure
used to study the two $\nu=2/3$ states had to be carefully designed
(with the 2DEG confined in a narrow, 12 nm wide, quantum-well), as
we aimed to have the transition between the two states at a sufficiently
high carrier density (and magnetic field), corresponding to having
high mobility throughout the transition region in the phase space
between the two states. A conductive $\text{n}^{+}$ GaAs layer was
grown $\sim$1 $\mu$m below the 2DEG and served as a backgate, capable
of tuning the density from 1 to $2.5\times10^{11}\,\mathrm{cm^{-2}}$,
with a corresponding low temperature dark mobility of 1.5 to $3.5\times10^{6}\,\mathrm{cm^{2}V^{-1}s^{-1}}$.
Lock-in measurements were performed at $\sim$80 Hz with an input
current $I=1$ nA and an electron temperature of $\sim$35 mK (see
section 1of \citep{SuppMat} for additional fabrication information).}

\textcolor{black}{The evolution of the four-terminal longitudinal
$(R_{xx})$, and transverse resistance $(R_{xy})$, measured in a
$40\,\mu$m wide Hall-bar geometry, is plotted on Fig. 1A and C. As
reported previously \citep{Eisenstein1990,Smet2001,Kraus,Huels2004,Verdene2007},
a clear transition between the two-spin varieties of the $\nu=2/3$
states is visible in $R_{xx}$ (around $V_{\text{bg}}=-0.5$ V and
$B=10$ T in Fig. 1A. The finite $R_{xx}$ region corresponds to the
point where the system undergoes a first order quantum phase transition
between the spin unpolarized and the spin polarized $\nu=2/3$ state.
The transverse resistance $R_{xy}\simeq(2/3\,e^{2}/h)^{-1}\simeq38.7\,\mathrm{{k}\Omega}$,
however, remains constant on both side of the transition. As predicted
in \citep{MacDonald1990} the presence of an upstream current leads
to the deviation of the two-terminal resistance from the canonical
value $R_{\text{2t}}\simeq38.7\,\mathrm{{k}\Omega}$. We therefore
have conducted two-terminal resistance measurements of several samples,
consisting of two 60 $\mu$m-wide ohmic contacts separated by a distance
$L$ ranging from 4 to 15 $\mu$m. The large aspect ratio (of width
to length) minimizes backscattering between the propagating edge modes
on opposite sides of the mesa. As visible on Fig. 2A for $L=4$ $\mu$m,
in the ($B$, $V_{\mathrm{bg}}$) phase space corresponding to the
polarized state, we find a two-terminal resistance $R_{\mathrm{2t}}^{\uparrow\uparrow}\left(L=4\right)\simeq38.6\pm0.1\,\mathrm{{k}\Omega}$.
However, for the unpolarized state, the resistance plateau is found
to deviate from the quantized value showing $R_{\mathrm{2t}}^{\uparrow\downarrow}\left(L=4\right)\simeq35.5\,\mathrm{{k}\Omega}$.}

\textcolor{black}{}
\begin{figure}
\centering{}\textcolor{black}{\includegraphics[width=0.9\columnwidth]{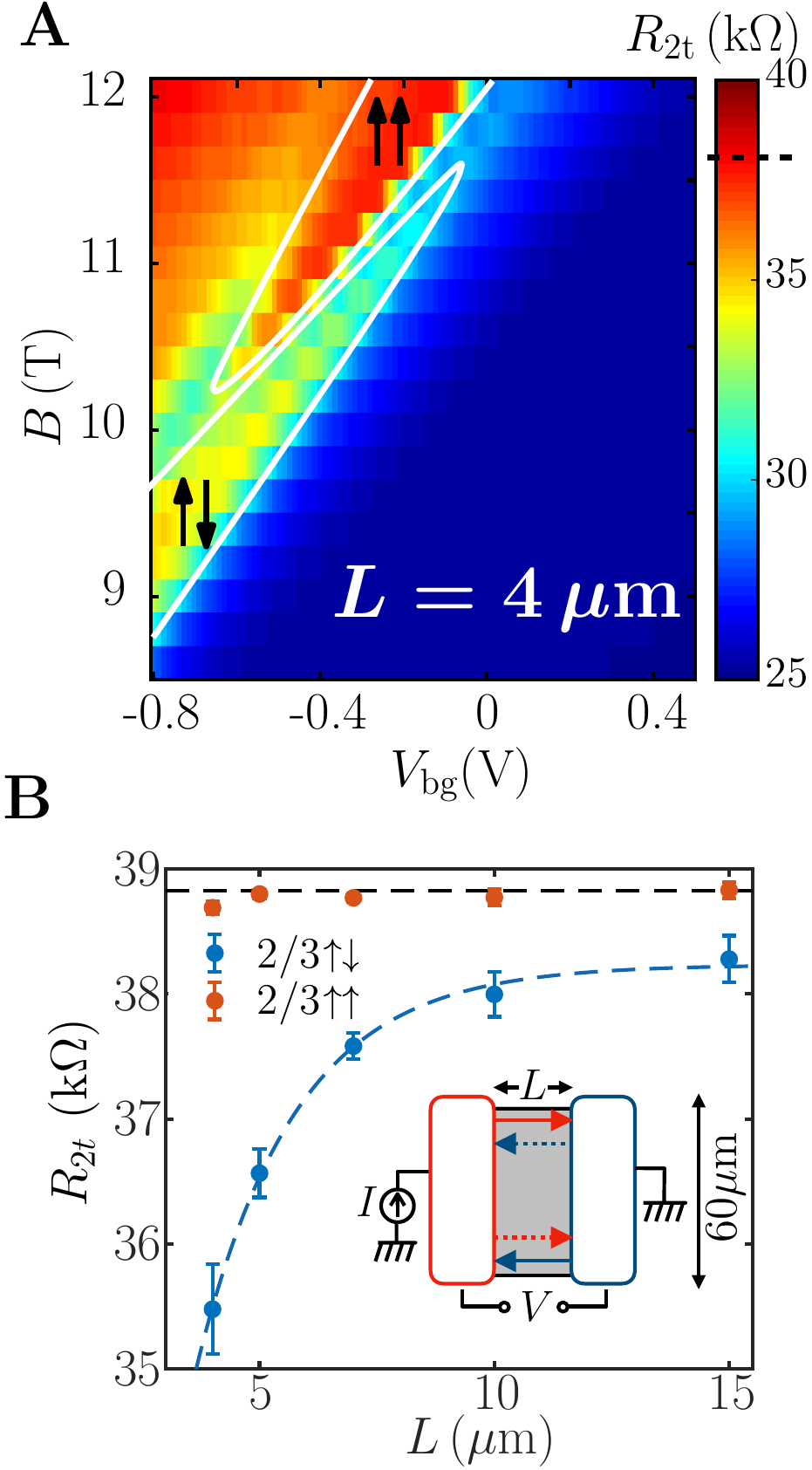}\caption{\textbf{Deviation from the quantized Hall resistance value owing to
the upstream current.} \textbf{(A)} Two-terminal magneto-resistance
versus backgate voltage for a $L=4\,\mu\text{m}$ long and $60\,\text{\ensuremath{\mu}m}$
wide sample measured at $T\sim35$ mK and $I=1$ nA. A clear difference
appears between the spin-polarized state, which remains quantized,
and the spin unpolarized state, which deviates substantially from
the quantized value. \textbf{(B)} Evolution of the two-terminal resistance
averaged over an area of $\left(B,V_{\text{bg}}\right)$ corresponding
to the polarized and unpolarized $\nu=2/3$ states, as a function
of the length $L$ \textcolor{black}{(see section 2 of \citep{SuppMat})}.
Dashed black line is the quantized value $\left(2/3\,e^{2}/h\right)^{-1}$
and dashed blue line is an exponential fit $R(x)=\left(R\left(0\right)-R\left(\infty\right)\right)e^{-x/l_{0}}+R\left(\infty\right)$,
where $R\left(0\right)=20\pm13\,\text{k}\Omega$, $R\left(\infty\right)=38.2\pm0.3\,\text{k}\Omega$
and $l_{0}=2.1\pm0.8\,\mu\text{m}$}
}
\end{figure}

\begin{figure*}
\noindent \centering{}\textcolor{black}{\includegraphics[width=1\textwidth]{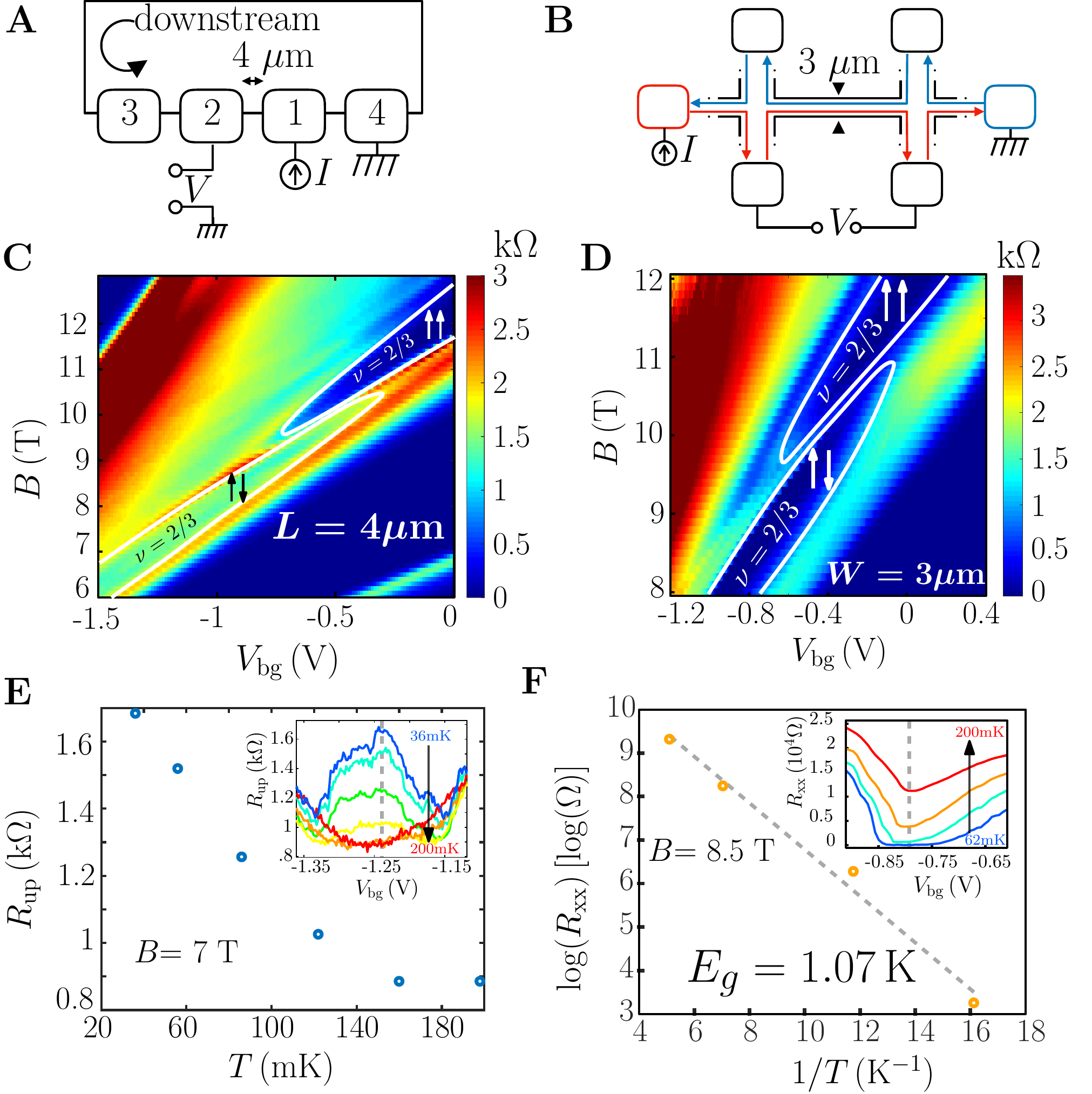}}\caption{\textbf{Different sample geometries validating the presence of a counter
propagating charge flow.} \textbf{(A)} Sketch of the 3-terminal measurement.
Four contacts are aligned on a single edge of the sample 4 $\mu$m
apart. Current $I=1$ nA was sourced at contact 1 and voltage was
measured between the contact 2 placed upstream and the ground; contact
4 was grounded and contact 3 was floating. \textbf{(B)} Sketch of
the 4-terminal $R_{xx}$ measurement on a narrow, 3 $\mu$m wide and
25 $\mu$m long Hall bar. The red lines represent the biased edge
channels whereas the blue ones represent the grounded edge channels.
\textbf{(C)} 3-terminal magnetoresistance versus the backgate voltage
for the measurement scheme presented in (A), a clear finite resistance
appears in the unpolarized region. \textbf{(D)} 4-terminal $R_{xx}$
versus backgate voltage for the measurement scheme presented in (B).
The $R_{xx}$ values are low in both polarized and unpolarized regions,
in contrast to (C). \textbf{(E)} Evolution of the resistance measured
in the unpolarized regime at $B=7\,\text{T}$ and $V_{\text{bg}}=-1.24\,V$
as function of temperature. Inset: Resistance in the unpolarized regime
as function of the backgate voltage for different temperatures (36,
56, 86, 122, 160 and 198 mK) . \textbf{(F)} Evolution of the log of
the longitudinal resistance versus the inverse of the temperature
in the Hall bar geometry. Inset: Evolution of the resistance versus
the backgate voltage in the unpolarized regime for several temperatures
(62, 85, 142 and 196 mK). The extracted activation gap is $E_{g}=1.07\,\text{K}$ }
\end{figure*}

\textcolor{black}{}
\begin{figure*}
\begin{centering}
\textcolor{black}{\includegraphics[width=1\textwidth]{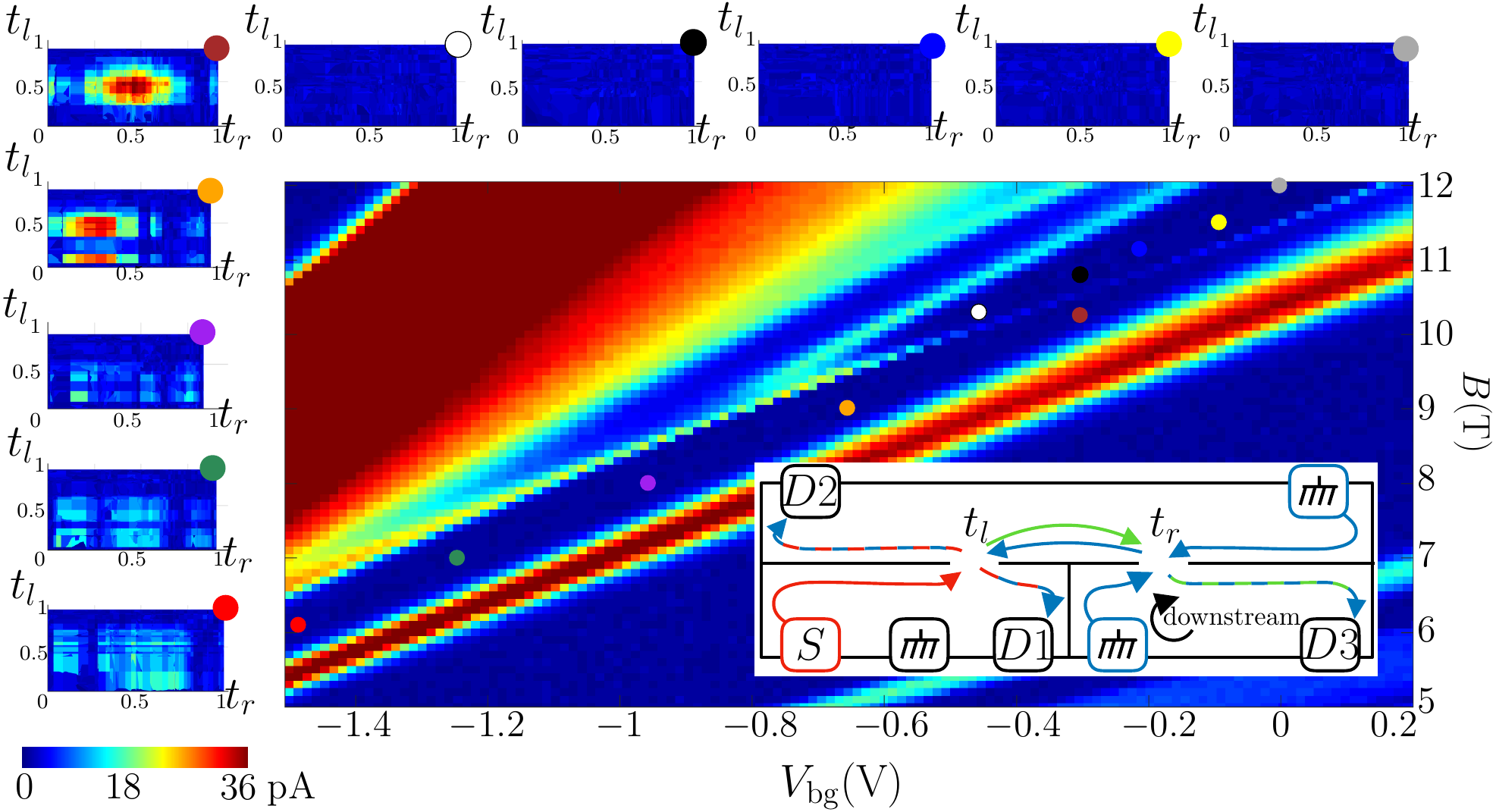}}
\par\end{centering}
\textcolor{black}{\caption{\textbf{Generation of upstream charge current by a quantum point contact.}
Evolution of the current measured in $D3$ as function of $t_{l}$
and $t_{r}$ for different points in the $\left(V_{\text{bg}},B\right)$
phase space (points shown in the main panel). A significant current
is measurable in the unpolarized region (red, green, purple, orange
and brown points) whereas no signal was measurable in the polarized
region (white, black, blue, yellow and gray points). Inset: Two successive
quantum point contacts setup. Current is sourced at $S$ (red), flowing
downstream to the left QPC; there, it is split to downstream (red/blue)
and upstream (green) charge currents. The unequilibrated upstream
current reaches the second QPC and turns back to downstream (green/blue),
where we measure its voltage at $D3$. The sketch is not to scale,
the distance between the two QPC is 700 nm and the distance between
the QPC and the nearest ohmic contact is 30 $\mu$m.}
}
\end{figure*}
\textcolor{black}{Measuring the evolution of $R_{\mathrm{2t}}^{\uparrow\uparrow}$
and $R_{\mathrm{2t}}^{\uparrow\downarrow}$ with length we find $R_{\mathrm{2t}}^{\uparrow\uparrow}$
independent of contact separation (Fig. 2B, orange circles) whereas
$R_{\mathrm{2t}}^{\uparrow\downarrow}$ increases with $L$, approaching
the quantized value for $L$=15 $\mu$m (Fig. 2B, blue circles). Exponential
fit of the two-terminal resistance is presented in Fig. 2B (dashed
blue line), $R(x)=(R(0)-R(\infty)\exp(-x/l_{0})+R(\infty)$, where
}$R\left(0\right)=20\pm13\,\text{k}\Omega$\textcolor{black}{{} is the
resistance at zero distance, }$R\left(\infty\right)=38.2\pm0.3\,\text{k}\Omega$\textcolor{black}{{}
is the resistance at infinite distance, }$l_{0}=2.1\pm0.8\,\mu\text{m}$\textcolor{black}{{}
is the characteristic equilibration length (see section 2 of \citep{SuppMat}
for additional details on Fig. 2B). Moreover, it is worth noting that
the resistance $R(0)$ is in agreement with the two-terminal resistance
predicted for unequilibrated channels proposed in \citep{MacDonald1990},
$R_{\text{2t}}=\left(4/3\,e^{2}/h\right)^{-1}\approx19.4\,\text{k}\Omega$.
These observation might have been possible at short distance due to
the reduction of scattering events and the screening of the Coulomb
interraction by the back gate placed $1\,\mu$m away from the 2DEG.}

\textcolor{black}{Bearing in mind that a finite $R_{xx}$ caused by
dissipation processes at short contact separation can lead to similar
observations, a few additional configurations were tested. One of
them was a configuration that employs a complementary Hall bar structure,
with a narrow Hall channel width of $3\,\mu$m (Fig. 3B); this was
necessary in order to ensure the lack of backscattering along distances
under consideration. The measured $R_{xx}$ (Fig. 3D) for the two
$\nu=2/3$ states using this geometry was negligibly small ensuring
that edge states located on opposite sides of the Hall bar (red and
blue lines on Fig. 3B) do not exchange particles. This observation
is in agreement with the relatively large gap ($\sim$1 K for the
unpolarized state) extracted from the temperature evolution of $R_{\mathrm{xx}}^{4p}$
presented on Fig. 3F, and in agreement with previous measurements
\citep{Engel1992,Boebinger1985}. Furthermore, testing a Corbino geometry
sample ensured a negligible bulk conductance of both $\nu=2/3$ states
(see section 3 of \citep{SuppMat}). Finally, a three-terminal configuration,
with contacts aligned on a single edge of the mesa (each separated
by 4 $\mu$m), allowed to separate the upstream current from the downstream
one (Fig. 3A), providing a direct measurement for the upstream conductance.
Current $I$ was sourced via contact 1 and drained via contact 4 to
the ground. A finite voltage $V_{\text{up}}$ was measured at contact
2 for the unpolarized state only as visible on Fig. 3C. The resistance,
defined as $R_{\text{up}}=V_{\text{up}}/I$ , continuously dropped
with increasing temperature up to 200 mK (Fig. 3E). This dependence
has an opposite trend to that of usual dissipative processes such
as variable range hopping or activation mechanisms, ruling them out
as alternative explanations. A complementary measurement of the downstream
resistance, $R_{\text{d}}=V_{\text{d}}/I$, was done by sourcing current
via contact 2 and measuring the voltage at contact 1. The upstream
and downstream conductances, calculated using the Landauer - Buttiker
formalism \citep{Buttiker1986,Datta1995} (see section 4 of \citep{SuppMat}),
leads to $G_{\text{d}}\approx0.687\,e^{2}/h$ and $G_{\text{up}}\approx0.026\,e^{2}/h$
or equivalent to a two-terminal resistance $\left(G_{\text{up}}+G_{\text{d}}\right)^{-1}\approx1.40\ h/e^{2}\approx36.2\,\mathrm{k}\Omega$,
in agreement with the two-terminal configuration at 4 $\mu$m presented
above (in Fig. 2). The mobility of the 2DEG in proximity to an alloyed
ohmic contact is degraded and its density is increased, which limited
us on the minimal distance between ohmic contacts to 4 $\mu$m. In
order to probe the edge modes at shorter distances we employed a configuration
consisting of two, gate defined quantum point contacts (QPCs) separated
by 700 nm shown in Fig. 4, inset, with all ohmic contacts placed far
away (above $30\mu\text{m}$). A current $I=1$ nA was sourced via
contact $S$; currents were monitored at the drains while scanning
the transmissions of the left and right QPCs $t_{l}$ and $t_{r}$.
This was done at different points in the ($V_{bg}$, $B$) phase space
for both spin polarization of the $\nu=2/3$ states, indicated by
the colored circles in Fig. 4. In the polarized state, all of the
current flowed to drains $D1$ and $D2$ independent of $t_{r}$,
consistent with downstream channels and zero current was measured
at $D3$ }(white, black, blue, yellow and gray points in Fig. 4)\textcolor{black}{.
However, in the unpolarized state }(red, green, purple, orange and
brown points in Fig. 4)\textcolor{black}{, substantial signal was
found in $D3$, simultaneously decreasing the current measured in
$D1$ and $D_{2}$ result in overall current conservation (see section
5 of \citep{SuppMat}). This \textquoteleft upstream effect\textquoteright{}
can be explained by the appearance of an upstream current between
the two QPCs (green arrow in Fig. 4 inset), which emerges from the
left QPC, flows a short distance to the right QPC, and scatters back
to the downstream channel, finally arriving at D3. Interestingly,
a maximum current at $D3$ was measured when $t_{l}=t_{r}=0.5$ (a
toy model for this effect is presented in Section 6 of \citep{SuppMat}).}

\textcolor{black}{The present set of experiments revealed counter-propagation
of charged particles in the fractional quantum Hall effect regime.
This present experiment may induce future theoretical works of the
less understood unpolarized $\nu=2/3$ state. }

\textbf{\textcolor{black}{Acknowledgements:}}\textcolor{black}{{} We
thank Ady Stern and Yigal Meir for fruitful discussions. We would
like to thank Diana Mahalu for her precious help in the ebeam lithography
process. }\textbf{\textcolor{black}{Funding:}}\textcolor{black}{{} We
acknowledge the European Research Council under the European Community\textquoteright s
Seventh Framework Program, grant agreement number 339070, the partial
support of the Minerva Foundation, grant number 711752, and, together
with V.U., the German Israeli Foundation (GIF), grant number I-1241-303.10/2014,
and the Israeli Science Foundation (ISF). }\textbf{\textcolor{black}{Author
contributions: }}\textcolor{black}{F.L. and A.R. contributed equally
to this work in sample design, device fabrication, measurement set-up,
data acquisition, data analysis and interpretation, and writing of
the paper. M.H. guided the experimental work and contributed in data
interpretation and writing of the paper. V.U. contributed in molecular
beam epitaxy growth. }\textbf{\textcolor{black}{Competing interests:
}}\textcolor{black}{The authors declare that they have no competing
financial interests. }\textbf{\textcolor{black}{Data and materials
availability: }}\textcolor{black}{All data needed to evaluate the
conclusions in the paper are present in the paper or the supplementary
materials. Additional data will be provided upon reasonable request
to the corresponding author.}

\bibliographystyle{apsrev4-1}

\begin{thebibliography}{42}%
\makeatletter
\providecommand \@ifxundefined [1]{%
 \@ifx{#1\undefined}
}%
\providecommand \@ifnum [1]{%
 \ifnum #1\expandafter \@firstoftwo
 \else \expandafter \@secondoftwo
 \fi
}%
\providecommand \@ifx [1]{%
 \ifx #1\expandafter \@firstoftwo
 \else \expandafter \@secondoftwo
 \fi
}%
\providecommand \natexlab [1]{#1}%
\providecommand \enquote  [1]{``#1''}%
\providecommand \bibnamefont  [1]{#1}%
\providecommand \bibfnamefont [1]{#1}%
\providecommand \citenamefont [1]{#1}%
\providecommand \href@noop [0]{\@secondoftwo}%
\providecommand \href [0]{\begingroup \@sanitize@url \@href}%
\providecommand \@href[1]{\@@startlink{#1}\@@href}%
\providecommand \@@href[1]{\endgroup#1\@@endlink}%
\providecommand \@sanitize@url [0]{\catcode `\\12\catcode `\$12\catcode
  `\&12\catcode `\#12\catcode `\^12\catcode `\_12\catcode `\%12\relax}%
\providecommand \@@startlink[1]{}%
\providecommand \@@endlink[0]{}%
\providecommand \url  [0]{\begingroup\@sanitize@url \@url }%
\providecommand \@url [1]{\endgroup\@href {#1}{\urlprefix }}%
\providecommand \urlprefix  [0]{URL }%
\providecommand \Eprint [0]{\href }%
\providecommand \doibase [0]{http://dx.doi.org/}%
\providecommand \selectlanguage [0]{\@gobble}%
\providecommand \bibinfo  [0]{\@secondoftwo}%
\providecommand \bibfield  [0]{\@secondoftwo}%
\providecommand \translation [1]{[#1]}%
\providecommand \BibitemOpen [0]{}%
\providecommand \bibitemStop [0]{}%
\providecommand \bibitemNoStop [0]{.\EOS\space}%
\providecommand \EOS [0]{\spacefactor3000\relax}%
\providecommand \BibitemShut  [1]{\csname bibitem#1\endcsname}%
\let\auto@bib@innerbib\@empty
\bibitem [{\citenamefont {Klitzing}\ \emph {et~al.}(1980)\citenamefont
  {Klitzing}, \citenamefont {Dorda},\ and\ \citenamefont
  {Pepper}}]{Klitzing1980}%
  \BibitemOpen
  \bibfield  {author} {\bibinfo {author} {\bibfnamefont {K.~V.}\ \bibnamefont
  {Klitzing}}, \bibinfo {author} {\bibfnamefont {G.}~\bibnamefont {Dorda}}, \
  and\ \bibinfo {author} {\bibfnamefont {M.}~\bibnamefont {Pepper}},\ }\href
  {\doibase 10.1103/PhysRevLett.45.494} {\bibfield  {journal} {\bibinfo
  {journal} {Physical Review Letters}\ }\textbf {\bibinfo {volume} {45}},\
  \bibinfo {pages} {494} (\bibinfo {year} {1980})}\BibitemShut {NoStop}%
\bibitem [{\citenamefont {Tsui}\ \emph {et~al.}(1982)\citenamefont {Tsui},
  \citenamefont {Stormer},\ and\ \citenamefont {Gossard}}]{Tsui1982}%
  \BibitemOpen
  \bibfield  {author} {\bibinfo {author} {\bibfnamefont {D.~C.}\ \bibnamefont
  {Tsui}}, \bibinfo {author} {\bibfnamefont {H.~L.}\ \bibnamefont {Stormer}}, \
  and\ \bibinfo {author} {\bibfnamefont {A.~C.}\ \bibnamefont {Gossard}},\
  }\href {\doibase 10.1103/PhysRevLett.48.1559} {\bibfield  {journal} {\bibinfo
   {journal} {Physical Review Letters}\ }\textbf {\bibinfo {volume} {48}},\
  \bibinfo {pages} {1559} (\bibinfo {year} {1982})}\BibitemShut {NoStop}%
\bibitem [{\citenamefont {Laughlin}(1983)}]{Laughlin1983}%
  \BibitemOpen
  \bibfield  {author} {\bibinfo {author} {\bibfnamefont {R.~B.}\ \bibnamefont
  {Laughlin}},\ }\href {\doibase 10.1103/PhysRevLett.50.1395} {\bibfield
  {journal} {\bibinfo  {journal} {Physical Review Letters}\ }\textbf {\bibinfo
  {volume} {50}},\ \bibinfo {pages} {1395} (\bibinfo {year}
  {1983})}\BibitemShut {NoStop}%
\bibitem [{\citenamefont {Beenakker}(1990)}]{Beenakker1990}%
  \BibitemOpen
  \bibfield  {author} {\bibinfo {author} {\bibfnamefont {C.~W.~J.}\
  \bibnamefont {Beenakker}},\ }\href {\doibase 10.1103/PhysRevLett.64.216}
  {\bibfield  {journal} {\bibinfo  {journal} {Physical Review Letters}\
  }\textbf {\bibinfo {volume} {64}},\ \bibinfo {pages} {216} (\bibinfo {year}
  {1990})}\BibitemShut {NoStop}%
\bibitem [{\citenamefont {Wen}(1991)}]{WEN1991}%
  \BibitemOpen
  \bibfield  {author} {\bibinfo {author} {\bibfnamefont {X.}~\bibnamefont
  {Wen}},\ }\href {\doibase 10.1142/S0217984991000058} {\bibfield  {journal}
  {\bibinfo  {journal} {Modern Physics Letters B}\ }\textbf {\bibinfo {volume}
  {05}},\ \bibinfo {pages} {39} (\bibinfo {year} {1991})}\BibitemShut {NoStop}%
\bibitem [{\citenamefont {MacDonald}(1990)}]{MacDonald1990}%
  \BibitemOpen
  \bibfield  {author} {\bibinfo {author} {\bibfnamefont {A.~H.}\ \bibnamefont
  {MacDonald}},\ }\href {\doibase 10.1103/PhysRevLett.64.220} {\bibfield
  {journal} {\bibinfo  {journal} {Physical Review Letters}\ }\textbf {\bibinfo
  {volume} {64}},\ \bibinfo {pages} {220} (\bibinfo {year} {1990})}\BibitemShut
  {NoStop}%
\bibitem [{\citenamefont {Ashoori}\ \emph {et~al.}(1992)\citenamefont
  {Ashoori}, \citenamefont {Stormer}, \citenamefont {Pfeiffer}, \citenamefont
  {Baldwin},\ and\ \citenamefont {West}}]{Ashoori1992}%
  \BibitemOpen
  \bibfield  {author} {\bibinfo {author} {\bibfnamefont {R.~C.}\ \bibnamefont
  {Ashoori}}, \bibinfo {author} {\bibfnamefont {H.~L.}\ \bibnamefont
  {Stormer}}, \bibinfo {author} {\bibfnamefont {L.~N.}\ \bibnamefont
  {Pfeiffer}}, \bibinfo {author} {\bibfnamefont {K.~W.}\ \bibnamefont
  {Baldwin}}, \ and\ \bibinfo {author} {\bibfnamefont {K.}~\bibnamefont
  {West}},\ }\href {\doibase 10.1103/PhysRevB.45.3894} {\bibfield  {journal}
  {\bibinfo  {journal} {Physical Review B}\ }\textbf {\bibinfo {volume} {45}},\
  \bibinfo {pages} {3894} (\bibinfo {year} {1992})}\BibitemShut {NoStop}%
\bibitem [{\citenamefont {Sabo}\ \emph {et~al.}(2017)\citenamefont {Sabo},
  \citenamefont {Gurman}, \citenamefont {Rosenblatt}, \citenamefont {Lafont},
  \citenamefont {Banitt}, \citenamefont {Park}, \citenamefont {Heiblum},
  \citenamefont {Gefen}, \citenamefont {Umansky},\ and\ \citenamefont
  {Mahalu}}]{Sabo2017}%
  \BibitemOpen
  \bibfield  {author} {\bibinfo {author} {\bibfnamefont {R.}~\bibnamefont
  {Sabo}}, \bibinfo {author} {\bibfnamefont {I.}~\bibnamefont {Gurman}},
  \bibinfo {author} {\bibfnamefont {A.}~\bibnamefont {Rosenblatt}}, \bibinfo
  {author} {\bibfnamefont {F.}~\bibnamefont {Lafont}}, \bibinfo {author}
  {\bibfnamefont {D.}~\bibnamefont {Banitt}}, \bibinfo {author} {\bibfnamefont
  {J.}~\bibnamefont {Park}}, \bibinfo {author} {\bibfnamefont {M.}~\bibnamefont
  {Heiblum}}, \bibinfo {author} {\bibfnamefont {Y.}~\bibnamefont {Gefen}},
  \bibinfo {author} {\bibfnamefont {V.}~\bibnamefont {Umansky}}, \ and\
  \bibinfo {author} {\bibfnamefont {D.}~\bibnamefont {Mahalu}},\ }\href
  {\doibase 10.1038/nphys4010} {\bibfield  {journal} {\bibinfo  {journal}
  {Nature Physics}\ } (\bibinfo {year} {2017}),\ 10.1038/nphys4010}\BibitemShut
  {NoStop}%
\bibitem [{\citenamefont {Kane}\ \emph {et~al.}(1994)\citenamefont {Kane},
  \citenamefont {Fisher},\ and\ \citenamefont {Polchinski}}]{Kane1994}%
  \BibitemOpen
  \bibfield  {author} {\bibinfo {author} {\bibfnamefont {C.~L.}\ \bibnamefont
  {Kane}}, \bibinfo {author} {\bibfnamefont {M.~P.~A.}\ \bibnamefont {Fisher}},
  \ and\ \bibinfo {author} {\bibfnamefont {J.}~\bibnamefont {Polchinski}},\
  }\href {\doibase 10.1103/PhysRevLett.72.4129} {\bibfield  {journal} {\bibinfo
   {journal} {Physical Review Letters}\ }\textbf {\bibinfo {volume} {72}},\
  \bibinfo {pages} {4129} (\bibinfo {year} {1994})}\BibitemShut {NoStop}%
\bibitem [{\citenamefont {Meir}(1994)}]{Meir1994}%
  \BibitemOpen
  \bibfield  {author} {\bibinfo {author} {\bibfnamefont {Y.}~\bibnamefont
  {Meir}},\ }\href {\doibase 10.1103/PhysRevLett.72.2624} {\bibfield  {journal}
  {\bibinfo  {journal} {Physical Review Letters}\ }\textbf {\bibinfo {volume}
  {72}},\ \bibinfo {pages} {2624} (\bibinfo {year} {1994})}\BibitemShut
  {NoStop}%
\bibitem [{\citenamefont {Wang}\ \emph {et~al.}()\citenamefont {Wang},
  \citenamefont {Meir},\ and\ \citenamefont {Gefen}}]{Wang}%
  \BibitemOpen
  \bibfield  {author} {\bibinfo {author} {\bibfnamefont {J.}~\bibnamefont
  {Wang}}, \bibinfo {author} {\bibfnamefont {Y.}~\bibnamefont {Meir}}, \ and\
  \bibinfo {author} {\bibfnamefont {Y.}~\bibnamefont {Gefen}},\ }\href
  {\doibase 10.1103/PhysRevLett.111.246803} {\
  10.1103/PhysRevLett.111.246803}\BibitemShut {NoStop}%
\bibitem [{\citenamefont {Bid}\ \emph {et~al.}(2010)\citenamefont {Bid},
  \citenamefont {Ofek}, \citenamefont {Inoue}, \citenamefont {Heiblum},
  \citenamefont {Kane}, \citenamefont {Umansky},\ and\ \citenamefont
  {Mahalu}}]{Bid2010}%
  \BibitemOpen
  \bibfield  {author} {\bibinfo {author} {\bibfnamefont {A.}~\bibnamefont
  {Bid}}, \bibinfo {author} {\bibfnamefont {N.}~\bibnamefont {Ofek}}, \bibinfo
  {author} {\bibfnamefont {H.}~\bibnamefont {Inoue}}, \bibinfo {author}
  {\bibfnamefont {M.}~\bibnamefont {Heiblum}}, \bibinfo {author} {\bibfnamefont
  {C.~L.}\ \bibnamefont {Kane}}, \bibinfo {author} {\bibfnamefont
  {V.}~\bibnamefont {Umansky}}, \ and\ \bibinfo {author} {\bibfnamefont
  {D.}~\bibnamefont {Mahalu}},\ }\href {\doibase 10.1038/nature09277}
  {\bibfield  {journal} {\bibinfo  {journal} {Nature}\ }\textbf {\bibinfo
  {volume} {466}},\ \bibinfo {pages} {585} (\bibinfo {year}
  {2010})}\BibitemShut {NoStop}%
\bibitem [{\citenamefont {Gurman}\ \emph {et~al.}(2012)\citenamefont {Gurman},
  \citenamefont {Sabo}, \citenamefont {Heiblum}, \citenamefont {Umansky},\ and\
  \citenamefont {Mahalu}}]{Gurman2012}%
  \BibitemOpen
  \bibfield  {author} {\bibinfo {author} {\bibfnamefont {I.}~\bibnamefont
  {Gurman}}, \bibinfo {author} {\bibfnamefont {R.}~\bibnamefont {Sabo}},
  \bibinfo {author} {\bibfnamefont {M.}~\bibnamefont {Heiblum}}, \bibinfo
  {author} {\bibfnamefont {V.}~\bibnamefont {Umansky}}, \ and\ \bibinfo
  {author} {\bibfnamefont {D.}~\bibnamefont {Mahalu}},\ }\href {\doibase
  10.1038/ncomms2305} {\bibfield  {journal} {\bibinfo  {journal} {Nature
  Communications}\ }\textbf {\bibinfo {volume} {3}},\ \bibinfo {pages} {1289}
  (\bibinfo {year} {2012})}\BibitemShut {NoStop}%
\bibitem [{\citenamefont {Inoue}\ \emph {et~al.}(2014)\citenamefont {Inoue},
  \citenamefont {Grivnin}, \citenamefont {Ronen}, \citenamefont {Heiblum},
  \citenamefont {Umansky},\ and\ \citenamefont {Mahalu}}]{Inoue2014a}%
  \BibitemOpen
  \bibfield  {author} {\bibinfo {author} {\bibfnamefont {H.}~\bibnamefont
  {Inoue}}, \bibinfo {author} {\bibfnamefont {A.}~\bibnamefont {Grivnin}},
  \bibinfo {author} {\bibfnamefont {Y.}~\bibnamefont {Ronen}}, \bibinfo
  {author} {\bibfnamefont {M.}~\bibnamefont {Heiblum}}, \bibinfo {author}
  {\bibfnamefont {V.}~\bibnamefont {Umansky}}, \ and\ \bibinfo {author}
  {\bibfnamefont {D.}~\bibnamefont {Mahalu}},\ }\href {\doibase
  10.1038/ncomms5067} {\bibfield  {journal} {\bibinfo  {journal} {Nature
  Communications}\ }\textbf {\bibinfo {volume} {5}} (\bibinfo {year} {2014}),\
  10.1038/ncomms5067}\BibitemShut {NoStop}%
\bibitem [{\citenamefont {Rosenblatt}\ \emph {et~al.}(2017)\citenamefont
  {Rosenblatt}, \citenamefont {Lafont}, \citenamefont {Levkivskyi},
  \citenamefont {Sabo}, \citenamefont {Gurman}, \citenamefont {Banitt},
  \citenamefont {Heiblum},\ and\ \citenamefont
  {Umansky}}]{rosenblatt2017transmission}%
  \BibitemOpen
  \bibfield  {author} {\bibinfo {author} {\bibfnamefont {A.}~\bibnamefont
  {Rosenblatt}}, \bibinfo {author} {\bibfnamefont {F.}~\bibnamefont {Lafont}},
  \bibinfo {author} {\bibfnamefont {I.}~\bibnamefont {Levkivskyi}}, \bibinfo
  {author} {\bibfnamefont {R.}~\bibnamefont {Sabo}}, \bibinfo {author}
  {\bibfnamefont {I.}~\bibnamefont {Gurman}}, \bibinfo {author} {\bibfnamefont
  {D.}~\bibnamefont {Banitt}}, \bibinfo {author} {\bibfnamefont
  {M.}~\bibnamefont {Heiblum}}, \ and\ \bibinfo {author} {\bibfnamefont
  {V.}~\bibnamefont {Umansky}},\ }\href@noop {} {\bibfield  {journal} {\bibinfo
   {journal} {Nature communications}\ }\textbf {\bibinfo {volume} {8}},\
  \bibinfo {pages} {2251} (\bibinfo {year} {2017})}\BibitemShut {NoStop}%
\bibitem [{\citenamefont {Grivnin}\ \emph {et~al.}(2014)\citenamefont
  {Grivnin}, \citenamefont {Inoue}, \citenamefont {Ronen}, \citenamefont
  {Baum}, \citenamefont {Heiblum}, \citenamefont {Umansky},\ and\ \citenamefont
  {Mahalu}}]{Grivnin2014}%
  \BibitemOpen
  \bibfield  {author} {\bibinfo {author} {\bibfnamefont {A.}~\bibnamefont
  {Grivnin}}, \bibinfo {author} {\bibfnamefont {H.}~\bibnamefont {Inoue}},
  \bibinfo {author} {\bibfnamefont {Y.}~\bibnamefont {Ronen}}, \bibinfo
  {author} {\bibfnamefont {Y.}~\bibnamefont {Baum}}, \bibinfo {author}
  {\bibfnamefont {M.}~\bibnamefont {Heiblum}}, \bibinfo {author} {\bibfnamefont
  {V.}~\bibnamefont {Umansky}}, \ and\ \bibinfo {author} {\bibfnamefont
  {D.}~\bibnamefont {Mahalu}},\ }\href {\doibase
  10.1103/PhysRevLett.113.266803} {\bibfield  {journal} {\bibinfo  {journal}
  {Phys. Rev. Lett.}\ }\textbf {\bibinfo {volume} {113}},\ \bibinfo {pages}
  {266803} (\bibinfo {year} {2014})}\BibitemShut {NoStop}%
\bibitem [{\citenamefont {Eisenstein}\ \emph {et~al.}(1990)\citenamefont
  {Eisenstein}, \citenamefont {Stormer}, \citenamefont {Pfeiffer},\ and\
  \citenamefont {West}}]{Eisenstein1990}%
  \BibitemOpen
  \bibfield  {author} {\bibinfo {author} {\bibfnamefont {J.~P.}\ \bibnamefont
  {Eisenstein}}, \bibinfo {author} {\bibfnamefont {H.~L.}\ \bibnamefont
  {Stormer}}, \bibinfo {author} {\bibfnamefont {L.~N.}\ \bibnamefont
  {Pfeiffer}}, \ and\ \bibinfo {author} {\bibfnamefont {K.~W.}\ \bibnamefont
  {West}},\ }\href {\doibase 10.1103/PhysRevB.41.7910} {\bibfield  {journal}
  {\bibinfo  {journal} {Physical Review B}\ }\textbf {\bibinfo {volume} {41}},\
  \bibinfo {pages} {7910} (\bibinfo {year} {1990})}\BibitemShut {NoStop}%
\bibitem [{\citenamefont {Kronm{\"{u}}ller}\ \emph {et~al.}(1998)\citenamefont
  {Kronm{\"{u}}ller}, \citenamefont {Dietsche}, \citenamefont {Weis},
  \citenamefont {von Klitzing}, \citenamefont {Wegscheider},\ and\
  \citenamefont {Bichler}}]{Kronmuller1998}%
  \BibitemOpen
  \bibfield  {author} {\bibinfo {author} {\bibfnamefont {S.}~\bibnamefont
  {Kronm{\"{u}}ller}}, \bibinfo {author} {\bibfnamefont {W.}~\bibnamefont
  {Dietsche}}, \bibinfo {author} {\bibfnamefont {J.}~\bibnamefont {Weis}},
  \bibinfo {author} {\bibfnamefont {K.}~\bibnamefont {von Klitzing}}, \bibinfo
  {author} {\bibfnamefont {W.}~\bibnamefont {Wegscheider}}, \ and\ \bibinfo
  {author} {\bibfnamefont {M.}~\bibnamefont {Bichler}},\ }\href {\doibase
  10.1103/PhysRevLett.81.2526} {\bibfield  {journal} {\bibinfo  {journal}
  {Physical Review Letters}\ }\textbf {\bibinfo {volume} {81}},\ \bibinfo
  {pages} {2526} (\bibinfo {year} {1998})}\BibitemShut {NoStop}%
\bibitem [{\citenamefont {Kronm{\"{u}}ller}\ \emph {et~al.}(1999)\citenamefont
  {Kronm{\"{u}}ller}, \citenamefont {Dietsche}, \citenamefont {v.~Klitzing},
  \citenamefont {Denninger}, \citenamefont {Wegscheider},\ and\ \citenamefont
  {Bichler}}]{Kronmuller1999}%
  \BibitemOpen
  \bibfield  {author} {\bibinfo {author} {\bibfnamefont {S.}~\bibnamefont
  {Kronm{\"{u}}ller}}, \bibinfo {author} {\bibfnamefont {W.}~\bibnamefont
  {Dietsche}}, \bibinfo {author} {\bibfnamefont {K.}~\bibnamefont
  {v.~Klitzing}}, \bibinfo {author} {\bibfnamefont {G.}~\bibnamefont
  {Denninger}}, \bibinfo {author} {\bibfnamefont {W.}~\bibnamefont
  {Wegscheider}}, \ and\ \bibinfo {author} {\bibfnamefont {M.}~\bibnamefont
  {Bichler}},\ }\href {\doibase 10.1103/PhysRevLett.82.4070} {\bibfield
  {journal} {\bibinfo  {journal} {Physical Review Letters}\ }\textbf {\bibinfo
  {volume} {82}},\ \bibinfo {pages} {4070} (\bibinfo {year}
  {1999})}\BibitemShut {NoStop}%
\bibitem [{\citenamefont {Chakraborty}(2000)}]{Chakraborty2000}%
  \BibitemOpen
  \bibfield  {author} {\bibinfo {author} {\bibfnamefont {T.}~\bibnamefont
  {Chakraborty}},\ }\href {\doibase 10.1080/00018730050198161} {\bibfield
  {journal} {\bibinfo  {journal} {Advances in Physics}\ }\textbf {\bibinfo
  {volume} {49}},\ \bibinfo {pages} {959} (\bibinfo {year} {2000})}\BibitemShut
  {NoStop}%
\bibitem [{\citenamefont {Smet}\ \emph
  {et~al.}(2001{\natexlab{a}})\citenamefont {Smet}, \citenamefont
  {Deutschmann}, \citenamefont {Wegscheider}, \citenamefont {Abstreiter},\ and\
  \citenamefont {von Klitzing}}]{Smet2001b}%
  \BibitemOpen
  \bibfield  {author} {\bibinfo {author} {\bibfnamefont {J.~H.}\ \bibnamefont
  {Smet}}, \bibinfo {author} {\bibfnamefont {R.~A.}\ \bibnamefont
  {Deutschmann}}, \bibinfo {author} {\bibfnamefont {W.}~\bibnamefont
  {Wegscheider}}, \bibinfo {author} {\bibfnamefont {G.}~\bibnamefont
  {Abstreiter}}, \ and\ \bibinfo {author} {\bibfnamefont {K.}~\bibnamefont {von
  Klitzing}},\ }\href {\doibase 10.1103/PhysRevLett.86.2412} {\bibfield
  {journal} {\bibinfo  {journal} {Physical Review Letters}\ }\textbf {\bibinfo
  {volume} {86}},\ \bibinfo {pages} {2412} (\bibinfo {year}
  {2001}{\natexlab{a}})}\BibitemShut {NoStop}%
\bibitem [{\citenamefont {Kraus}\ \emph {et~al.}()\citenamefont {Kraus},
  \citenamefont {Stern}, \citenamefont {Lok}, \citenamefont {Dietsche},
  \citenamefont {{Von Klitzing}}, \citenamefont {Bichler}, \citenamefont
  {Schuh},\ and\ \citenamefont {Wegscheider}}]{Kraus}%
  \BibitemOpen
  \bibfield  {author} {\bibinfo {author} {\bibfnamefont {S.}~\bibnamefont
  {Kraus}}, \bibinfo {author} {\bibfnamefont {O.}~\bibnamefont {Stern}},
  \bibinfo {author} {\bibfnamefont {J.~G.~S.}\ \bibnamefont {Lok}}, \bibinfo
  {author} {\bibfnamefont {W.}~\bibnamefont {Dietsche}}, \bibinfo {author}
  {\bibfnamefont {K.}~\bibnamefont {{Von Klitzing}}}, \bibinfo {author}
  {\bibfnamefont {M.}~\bibnamefont {Bichler}}, \bibinfo {author} {\bibfnamefont
  {D.}~\bibnamefont {Schuh}}, \ and\ \bibinfo {author} {\bibfnamefont
  {W.}~\bibnamefont {Wegscheider}},\ }\href {\doibase
  10.1103/PhysRevLett.89.266801} {\ \textbf {\bibinfo {volume} {21}},\
  10.1103/PhysRevLett.89.266801}\BibitemShut {NoStop}%
\bibitem [{\citenamefont {Hayakawa}\ \emph {et~al.}(2012)\citenamefont
  {Hayakawa}, \citenamefont {Muraki},\ and\ \citenamefont
  {Yusa}}]{Hayakawa2012}%
  \BibitemOpen
  \bibfield  {author} {\bibinfo {author} {\bibfnamefont {J.}~\bibnamefont
  {Hayakawa}}, \bibinfo {author} {\bibfnamefont {K.}~\bibnamefont {Muraki}}, \
  and\ \bibinfo {author} {\bibfnamefont {G.}~\bibnamefont {Yusa}},\ }\href
  {\doibase 10.1038/nnano.2012.209} {\bibfield  {journal} {\bibinfo  {journal}
  {Nature Nanotechnology}\ }\textbf {\bibinfo {volume} {8}},\ \bibinfo {pages}
  {31} (\bibinfo {year} {2012})}\BibitemShut {NoStop}%
\bibitem [{\citenamefont {Moore}\ \emph {et~al.}(2017)\citenamefont {Moore},
  \citenamefont {Hayakawa}, \citenamefont {Mano}, \citenamefont {Noda},\ and\
  \citenamefont {Yusa}}]{Moore2017}%
  \BibitemOpen
  \bibfield  {author} {\bibinfo {author} {\bibfnamefont {J.~N.}\ \bibnamefont
  {Moore}}, \bibinfo {author} {\bibfnamefont {J.}~\bibnamefont {Hayakawa}},
  \bibinfo {author} {\bibfnamefont {T.}~\bibnamefont {Mano}}, \bibinfo {author}
  {\bibfnamefont {T.}~\bibnamefont {Noda}}, \ and\ \bibinfo {author}
  {\bibfnamefont {G.}~\bibnamefont {Yusa}},\ }\href {\doibase
  10.1103/PhysRevLett.118.076802} {\bibfield  {journal} {\bibinfo  {journal}
  {Phys. Rev. Lett.}\ }\textbf {\bibinfo {volume} {118}},\ \bibinfo {pages}
  {076802} (\bibinfo {year} {2017})}\BibitemShut {NoStop}%
\bibitem [{\citenamefont {Mong}\ \emph {et~al.}(2014)\citenamefont {Mong},
  \citenamefont {Clarke}, \citenamefont {Alicea}, \citenamefont {Lindner},
  \citenamefont {Fendley}, \citenamefont {Nayak}, \citenamefont {Oreg},
  \citenamefont {Stern}, \citenamefont {Berg}, \citenamefont {Shtengel},\ and\
  \citenamefont {Fisher}}]{Mong2014}%
  \BibitemOpen
  \bibfield  {author} {\bibinfo {author} {\bibfnamefont {R.~S.~K.}\
  \bibnamefont {Mong}}, \bibinfo {author} {\bibfnamefont {D.~J.}\ \bibnamefont
  {Clarke}}, \bibinfo {author} {\bibfnamefont {J.}~\bibnamefont {Alicea}},
  \bibinfo {author} {\bibfnamefont {N.~H.}\ \bibnamefont {Lindner}}, \bibinfo
  {author} {\bibfnamefont {P.}~\bibnamefont {Fendley}}, \bibinfo {author}
  {\bibfnamefont {C.}~\bibnamefont {Nayak}}, \bibinfo {author} {\bibfnamefont
  {Y.}~\bibnamefont {Oreg}}, \bibinfo {author} {\bibfnamefont {A.}~\bibnamefont
  {Stern}}, \bibinfo {author} {\bibfnamefont {E.}~\bibnamefont {Berg}},
  \bibinfo {author} {\bibfnamefont {K.}~\bibnamefont {Shtengel}}, \ and\
  \bibinfo {author} {\bibfnamefont {M.~P.~A.}\ \bibnamefont {Fisher}},\ }\href
  {\doibase 10.1103/PhysRevX.4.011036} {\bibfield  {journal} {\bibinfo
  {journal} {Physical Review X}\ }\textbf {\bibinfo {volume} {4}} (\bibinfo
  {year} {2014}),\ 10.1103/PhysRevX.4.011036},\ \Eprint
  {http://arxiv.org/abs/1307.4403} {arXiv:1307.4403} \BibitemShut {NoStop}%
\bibitem [{\citenamefont {Clarke}\ \emph {et~al.}(2014)\citenamefont {Clarke},
  \citenamefont {Alicea},\ and\ \citenamefont {Shtengel}}]{Clarke2014}%
  \BibitemOpen
  \bibfield  {author} {\bibinfo {author} {\bibfnamefont {D.~J.}\ \bibnamefont
  {Clarke}}, \bibinfo {author} {\bibfnamefont {J.}~\bibnamefont {Alicea}}, \
  and\ \bibinfo {author} {\bibfnamefont {K.}~\bibnamefont {Shtengel}},\ }\href
  {\doibase 10.1038/nphys3114} {\bibfield  {journal} {\bibinfo  {journal}
  {Nature Physics}\ }\textbf {\bibinfo {volume} {10}},\ \bibinfo {pages} {877}
  (\bibinfo {year} {2014})},\ \Eprint {http://arxiv.org/abs/1312.6123}
  {arXiv:1312.6123} \BibitemShut {NoStop}%
\bibitem [{\citenamefont {Wu}\ \emph {et~al.}(2018)\citenamefont {Wu},
  \citenamefont {Wan}, \citenamefont {Kazakov}, \citenamefont {Wang},
  \citenamefont {Simion}, \citenamefont {Liang}, \citenamefont {West},
  \citenamefont {Baldwin}, \citenamefont {Pfeiffer}, \citenamefont
  {Lyanda-Geller},\ and\ \citenamefont {Rokhinson}}]{Wu2018}%
  \BibitemOpen
  \bibfield  {author} {\bibinfo {author} {\bibfnamefont {T.}~\bibnamefont
  {Wu}}, \bibinfo {author} {\bibfnamefont {Z.}~\bibnamefont {Wan}}, \bibinfo
  {author} {\bibfnamefont {A.}~\bibnamefont {Kazakov}}, \bibinfo {author}
  {\bibfnamefont {Y.}~\bibnamefont {Wang}}, \bibinfo {author} {\bibfnamefont
  {G.}~\bibnamefont {Simion}}, \bibinfo {author} {\bibfnamefont
  {J.}~\bibnamefont {Liang}}, \bibinfo {author} {\bibfnamefont {K.~W.}\
  \bibnamefont {West}}, \bibinfo {author} {\bibfnamefont {K.}~\bibnamefont
  {Baldwin}}, \bibinfo {author} {\bibfnamefont {L.~N.}\ \bibnamefont
  {Pfeiffer}}, \bibinfo {author} {\bibfnamefont {Y.}~\bibnamefont
  {Lyanda-Geller}}, \ and\ \bibinfo {author} {\bibfnamefont {L.~P.}\
  \bibnamefont {Rokhinson}},\ }\href {\doibase 10.1103/PhysRevB.97.245304}
  {\bibfield  {journal} {\bibinfo  {journal} {Phys. Rev. B}\ }\textbf {\bibinfo
  {volume} {97}},\ \bibinfo {pages} {245304} (\bibinfo {year}
  {2018})}\BibitemShut {NoStop}%
\bibitem [{\citenamefont {Jain}(1989)}]{Jain1989}%
  \BibitemOpen
  \bibfield  {author} {\bibinfo {author} {\bibfnamefont {J.~K.}\ \bibnamefont
  {Jain}},\ }\href {\doibase 10.1103/PhysRevLett.63.199} {\bibfield  {journal}
  {\bibinfo  {journal} {Physical Review Letters}\ }\textbf {\bibinfo {volume}
  {63}},\ \bibinfo {pages} {199} (\bibinfo {year} {1989})}\BibitemShut
  {NoStop}%
\bibitem [{\citenamefont {Kukushkin}\ \emph {et~al.}(1999)\citenamefont
  {Kukushkin}, \citenamefont {Klitzing},\ and\ \citenamefont
  {Eberl}}]{Kukushkin1999}%
  \BibitemOpen
  \bibfield  {author} {\bibinfo {author} {\bibfnamefont {I.~V.}\ \bibnamefont
  {Kukushkin}}, \bibinfo {author} {\bibfnamefont {K.~V.}\ \bibnamefont
  {Klitzing}}, \ and\ \bibinfo {author} {\bibfnamefont {K.}~\bibnamefont
  {Eberl}},\ }\href@noop {} {\ \textbf {\bibinfo {volume} {82}} (\bibinfo
  {year} {1999})}\BibitemShut {NoStop}%
\bibitem [{\citenamefont {Verdene}\ \emph {et~al.}(2007)\citenamefont
  {Verdene}, \citenamefont {Martin}, \citenamefont {Gamez}, \citenamefont
  {Smet}, \citenamefont {von Klitzing}, \citenamefont {Mahalu}, \citenamefont
  {Schuh}, \citenamefont {Abstreiter},\ and\ \citenamefont
  {Yacoby}}]{Verdene2007}%
  \BibitemOpen
  \bibfield  {author} {\bibinfo {author} {\bibfnamefont {B.}~\bibnamefont
  {Verdene}}, \bibinfo {author} {\bibfnamefont {J.}~\bibnamefont {Martin}},
  \bibinfo {author} {\bibfnamefont {G.}~\bibnamefont {Gamez}}, \bibinfo
  {author} {\bibfnamefont {J.}~\bibnamefont {Smet}}, \bibinfo {author}
  {\bibfnamefont {K.}~\bibnamefont {von Klitzing}}, \bibinfo {author}
  {\bibfnamefont {D.}~\bibnamefont {Mahalu}}, \bibinfo {author} {\bibfnamefont
  {D.}~\bibnamefont {Schuh}}, \bibinfo {author} {\bibfnamefont
  {G.}~\bibnamefont {Abstreiter}}, \ and\ \bibinfo {author} {\bibfnamefont
  {A.}~\bibnamefont {Yacoby}},\ }\href {\doibase 10.1038/nphys588} {\bibfield
  {journal} {\bibinfo  {journal} {Nature Physics}\ }\textbf {\bibinfo {volume}
  {3}},\ \bibinfo {pages} {392} (\bibinfo {year} {2007})}\BibitemShut {NoStop}%
\bibitem [{\citenamefont {Smet}\ \emph
  {et~al.}(2001{\natexlab{b}})\citenamefont {Smet}, \citenamefont
  {Deutschmann}, \citenamefont {Wegscheider}, \citenamefont {Abstreiter},\ and\
  \citenamefont {{Von Klitzing}}}]{Smet2001}%
  \BibitemOpen
  \bibfield  {author} {\bibinfo {author} {\bibfnamefont {J.~H.}\ \bibnamefont
  {Smet}}, \bibinfo {author} {\bibfnamefont {R.~A.}\ \bibnamefont
  {Deutschmann}}, \bibinfo {author} {\bibfnamefont {W.}~\bibnamefont
  {Wegscheider}}, \bibinfo {author} {\bibfnamefont {G.}~\bibnamefont
  {Abstreiter}}, \ and\ \bibinfo {author} {\bibfnamefont {K.}~\bibnamefont
  {{Von Klitzing}}},\ }\href {\doibase 10.1103/PhysRevLett.86.2412} {\bibfield
  {journal} {\bibinfo  {journal} {Physical Review Letters}\ }\textbf {\bibinfo
  {volume} {86}},\ \bibinfo {pages} {2412} (\bibinfo {year}
  {2001}{\natexlab{b}})}\BibitemShut {NoStop}%
\bibitem [{\citenamefont {Huels}\ \emph {et~al.}(2004)\citenamefont {Huels},
  \citenamefont {Weis}, \citenamefont {Smet}, \citenamefont {Klitzing},\ and\
  \citenamefont {Wasilewski}}]{Huels2004}%
  \BibitemOpen
  \bibfield  {author} {\bibinfo {author} {\bibfnamefont {J.}~\bibnamefont
  {Huels}}, \bibinfo {author} {\bibfnamefont {J.}~\bibnamefont {Weis}},
  \bibinfo {author} {\bibfnamefont {J.}~\bibnamefont {Smet}}, \bibinfo {author}
  {\bibfnamefont {K.}~\bibnamefont {Klitzing}}, \ and\ \bibinfo {author}
  {\bibfnamefont {Z.}~\bibnamefont {Wasilewski}},\ }\href {\doibase
  10.1103/PhysRevB.69.085319} {\bibfield  {journal} {\bibinfo  {journal}
  {Physical Review B}\ }\textbf {\bibinfo {volume} {69}},\ \bibinfo {pages}
  {085319} (\bibinfo {year} {2004})}\BibitemShut {NoStop}%
\bibitem [{\citenamefont {Hennel}\ \emph {et~al.}(2016)\citenamefont {Hennel},
  \citenamefont {Braem}, \citenamefont {Baer}, \citenamefont {Tiemann},
  \citenamefont {Sohi}, \citenamefont {Wehrli}, \citenamefont {Hofmann},
  \citenamefont {Reichl}, \citenamefont {Wegscheider}, \citenamefont
  {R\"ossler}, \citenamefont {Ihn}, \citenamefont {Ensslin}, \citenamefont
  {Rudner},\ and\ \citenamefont {Rosenow}}]{Hennel2016}%
  \BibitemOpen
  \bibfield  {author} {\bibinfo {author} {\bibfnamefont {S.}~\bibnamefont
  {Hennel}}, \bibinfo {author} {\bibfnamefont {B.~A.}\ \bibnamefont {Braem}},
  \bibinfo {author} {\bibfnamefont {S.}~\bibnamefont {Baer}}, \bibinfo {author}
  {\bibfnamefont {L.}~\bibnamefont {Tiemann}}, \bibinfo {author} {\bibfnamefont
  {P.}~\bibnamefont {Sohi}}, \bibinfo {author} {\bibfnamefont {D.}~\bibnamefont
  {Wehrli}}, \bibinfo {author} {\bibfnamefont {A.}~\bibnamefont {Hofmann}},
  \bibinfo {author} {\bibfnamefont {C.}~\bibnamefont {Reichl}}, \bibinfo
  {author} {\bibfnamefont {W.}~\bibnamefont {Wegscheider}}, \bibinfo {author}
  {\bibfnamefont {C.}~\bibnamefont {R\"ossler}}, \bibinfo {author}
  {\bibfnamefont {T.}~\bibnamefont {Ihn}}, \bibinfo {author} {\bibfnamefont
  {K.}~\bibnamefont {Ensslin}}, \bibinfo {author} {\bibfnamefont {M.~S.}\
  \bibnamefont {Rudner}}, \ and\ \bibinfo {author} {\bibfnamefont
  {B.}~\bibnamefont {Rosenow}},\ }\href {\doibase
  10.1103/PhysRevLett.116.136804} {\bibfield  {journal} {\bibinfo  {journal}
  {Phys. Rev. Lett.}\ }\textbf {\bibinfo {volume} {116}},\ \bibinfo {pages}
  {136804} (\bibinfo {year} {2016})}\BibitemShut {NoStop}%
\bibitem [{\citenamefont {Cho}\ \emph {et~al.}(1998)\citenamefont {Cho},
  \citenamefont {Young}, \citenamefont {Kang}, \citenamefont {Campman},
  \citenamefont {Gossard}, \citenamefont {Bichler},\ and\ \citenamefont
  {Wegscheider}}]{Cho1998}%
  \BibitemOpen
  \bibfield  {author} {\bibinfo {author} {\bibfnamefont {H.}~\bibnamefont
  {Cho}}, \bibinfo {author} {\bibfnamefont {J.~B.}\ \bibnamefont {Young}},
  \bibinfo {author} {\bibfnamefont {W.}~\bibnamefont {Kang}}, \bibinfo {author}
  {\bibfnamefont {K.~L.}\ \bibnamefont {Campman}}, \bibinfo {author}
  {\bibfnamefont {A.~C.}\ \bibnamefont {Gossard}}, \bibinfo {author}
  {\bibfnamefont {M.}~\bibnamefont {Bichler}}, \ and\ \bibinfo {author}
  {\bibfnamefont {W.}~\bibnamefont {Wegscheider}},\ }\href {\doibase
  10.1103/PhysRevLett.81.2522} {\bibfield  {journal} {\bibinfo  {journal}
  {Phys. Rev. Lett.}\ }\textbf {\bibinfo {volume} {81}},\ \bibinfo {pages}
  {2522} (\bibinfo {year} {1998})}\BibitemShut {NoStop}%
\bibitem [{\citenamefont {Li}\ \emph {et~al.}(2012)\citenamefont {Li},
  \citenamefont {Umansky}, \citenamefont {von Klitzing},\ and\ \citenamefont
  {Smet}}]{Li2012}%
  \BibitemOpen
  \bibfield  {author} {\bibinfo {author} {\bibfnamefont {Y.~Q.}\ \bibnamefont
  {Li}}, \bibinfo {author} {\bibfnamefont {V.}~\bibnamefont {Umansky}},
  \bibinfo {author} {\bibfnamefont {K.}~\bibnamefont {von Klitzing}}, \ and\
  \bibinfo {author} {\bibfnamefont {J.~H.}\ \bibnamefont {Smet}},\ }\href
  {\doibase 10.1103/PhysRevB.86.115421} {\bibfield  {journal} {\bibinfo
  {journal} {Phys. Rev. B}\ }\textbf {\bibinfo {volume} {86}},\ \bibinfo
  {pages} {115421} (\bibinfo {year} {2012})}\BibitemShut {NoStop}%
\bibitem [{\citenamefont {Smet}\ \emph {et~al.}(2002)\citenamefont {Smet},
  \citenamefont {Deutschmann}, \citenamefont {Ertl}, \citenamefont
  {Wegscheider}, \citenamefont {Abstreiter},\ and\ \citenamefont {von
  Klitzing}}]{smet2002}%
  \BibitemOpen
  \bibfield  {author} {\bibinfo {author} {\bibfnamefont {J.~H.}\ \bibnamefont
  {Smet}}, \bibinfo {author} {\bibfnamefont {R.~A.}\ \bibnamefont
  {Deutschmann}}, \bibinfo {author} {\bibfnamefont {F.}~\bibnamefont {Ertl}},
  \bibinfo {author} {\bibfnamefont {W.}~\bibnamefont {Wegscheider}}, \bibinfo
  {author} {\bibfnamefont {G.}~\bibnamefont {Abstreiter}}, \ and\ \bibinfo
  {author} {\bibfnamefont {K.}~\bibnamefont {von Klitzing}},\ }\href
  {http://dx.doi.org/10.1038/415281a} {\bibfield  {journal} {\bibinfo
  {journal} {Nature}\ }\textbf {\bibinfo {volume} {415}},\ \bibinfo {pages}
  {281} (\bibinfo {year} {2002})}\BibitemShut {NoStop}%
\bibitem [{\citenamefont {Wu}\ \emph {et~al.}(2012)\citenamefont {Wu},
  \citenamefont {Sreejith},\ and\ \citenamefont {Jain}}]{Wu2012}%
  \BibitemOpen
  \bibfield  {author} {\bibinfo {author} {\bibfnamefont {Y.-H.}\ \bibnamefont
  {Wu}}, \bibinfo {author} {\bibfnamefont {G.~J.}\ \bibnamefont {Sreejith}}, \
  and\ \bibinfo {author} {\bibfnamefont {J.~K.}\ \bibnamefont {Jain}},\ }\href
  {\doibase 10.1103/PhysRevB.86.115127} {\bibfield  {journal} {\bibinfo
  {journal} {Phys. Rev. B}\ }\textbf {\bibinfo {volume} {86}},\ \bibinfo
  {pages} {115127} (\bibinfo {year} {2012})}\BibitemShut {NoStop}%
\bibitem [{Sup()}]{SuppMat}%
  \BibitemOpen
  \href@noop {} {\emph {\bibinfo {title} {Supplementary
  Materials}}}\BibitemShut {NoStop}%
\bibitem [{\citenamefont {Engel}\ \emph {et~al.}(1992)\citenamefont {Engel},
  \citenamefont {Hwang}, \citenamefont {Sajoto}, \citenamefont {Tsui},\ and\
  \citenamefont {Shayegan}}]{Engel1992}%
  \BibitemOpen
  \bibfield  {author} {\bibinfo {author} {\bibfnamefont {L.~W.}\ \bibnamefont
  {Engel}}, \bibinfo {author} {\bibfnamefont {S.~W.}\ \bibnamefont {Hwang}},
  \bibinfo {author} {\bibfnamefont {T.}~\bibnamefont {Sajoto}}, \bibinfo
  {author} {\bibfnamefont {D.~C.}\ \bibnamefont {Tsui}}, \ and\ \bibinfo
  {author} {\bibfnamefont {M.}~\bibnamefont {Shayegan}},\ }\href {\doibase
  10.1103/PhysRevB.45.3418} {\bibfield  {journal} {\bibinfo  {journal}
  {Physical Review B}\ }\textbf {\bibinfo {volume} {45}},\ \bibinfo {pages}
  {3418} (\bibinfo {year} {1992})}\BibitemShut {NoStop}%
\bibitem [{\citenamefont {Boebinger}\ \emph {et~al.}(1985)\citenamefont
  {Boebinger}, \citenamefont {Chang}, \citenamefont {Stormer},\ and\
  \citenamefont {Tsui}}]{Boebinger1985}%
  \BibitemOpen
  \bibfield  {author} {\bibinfo {author} {\bibfnamefont {G.~S.}\ \bibnamefont
  {Boebinger}}, \bibinfo {author} {\bibfnamefont {A.~M.}\ \bibnamefont
  {Chang}}, \bibinfo {author} {\bibfnamefont {H.~L.}\ \bibnamefont {Stormer}},
  \ and\ \bibinfo {author} {\bibfnamefont {D.~C.}\ \bibnamefont {Tsui}},\
  }\href {\doibase 10.1103/PhysRevLett.55.1606} {\bibfield  {journal} {\bibinfo
   {journal} {Physical Review Letters}\ }\textbf {\bibinfo {volume} {55}},\
  \bibinfo {pages} {1606} (\bibinfo {year} {1985})}\BibitemShut {NoStop}%
\bibitem [{\citenamefont {B{\"{u}}ttiker}(1986)}]{Buttiker1986}%
  \BibitemOpen
  \bibfield  {author} {\bibinfo {author} {\bibfnamefont {M.}~\bibnamefont
  {B{\"{u}}ttiker}},\ }\href {\doibase 10.1103/PhysRevLett.57.1761} {\bibfield
  {journal} {\bibinfo  {journal} {Physical Review Letters}\ }\textbf {\bibinfo
  {volume} {57}},\ \bibinfo {pages} {1761} (\bibinfo {year}
  {1986})}\BibitemShut {NoStop}%
\bibitem [{\citenamefont {Datta}(1995)}]{Datta1995}%
  \BibitemOpen
  \bibfield  {author} {\bibinfo {author} {\bibfnamefont {S.}~\bibnamefont
  {Datta}},\ }\href {\doibase 10.1063/1.2807624} {\emph {\bibinfo {title}
  {Cambridge University Press}}},\ Vol.~\bibinfo {volume} {3}\ (\bibinfo {year}
  {1995})\ p.\ \bibinfo {pages} {377}\BibitemShut {NoStop}%
\end{thebibliography}

\end{document}


\textcolor{black}{}

\textcolor{black}{}

\preprint{\textcolor{black}{}}

\title{\textcolor{black}{Counter-propagating charge transport in the quantum
Hall effect regime}\\
\textcolor{black}{{} Supplementary text}}

\author{\textcolor{black}{Fabien Lafont$^{1,2,*},$Amir Rosenblatt$^{1,*},$
Moty Heiblum$^{1},$ Vladimir Umansky$^{1}$}}

\affiliation{\textcolor{black}{$^{1}$Braun Center for Submicron Research, Dept.
of Condensed Matter physics, Weizmann Institute of Science, Rehovot
76100, Israel}}

\affiliation{\textcolor{black}{$^{2}$Collège de France, 11 place Marcelin Berthelot,
75231 Paris Cedex 05, France}}

\address{\textcolor{black}{$^{*}$These authors contributed equally to this
work}}
\maketitle

\subsubsection{\textcolor{black}{Sample fabrication }}

\textcolor{black}{A narrow quantum well (12nm) is buried 83 nm bellow
the surface of a GaAs-AlGaAs heterostructure. A conductive $\text{n}^{+}$
layer of GaAs was grown $\sim1$$\mu\text{m}$ below the 2DEG and
served as a back-gate. A window etched $\sim0.9$$\mu\text{m}$ deep,
followed by evaporation of 110 nm Au, 55 nm Ge and 40 nm Ni alloyed
to the heterostructure, give us a connection to the back-gate. The
same evaporation, on top of the heterostructure, also produce the
ohmic contacts. The QPCs were defined by split metallic gates consisting
of 5 nm Ti and 15 nm Au evaporated on the GaAs-AlGaAs heterostructure.
The opening of each QPC is 500 nm wide.}

\subsubsection{Two-terminal measurements}

Two-terminal magneto-resistance measurements were made on 60 $\mu\text{m}$
wide mesa contacted by two ohmic contacts separated by a distance
$L$ ranging from 4 to 15 $\mu$m. Current $I=1\text{nA}$ was sourced
in one contact and drained to ground from the other contact while
their voltage difference was monitored. Quantum Hall edge states are
formed at the edges of the sample as sketched in Fig. S1A: solid arrows
illustrate downstream (clockwise) chiral modes and dashed arrows illustrate
the upstream charge transport on the edge. Two-terminal resistance
$R_{\text{2t}}=V/I$ were measured as function of magnetic field and
backgate voltage on five separate devices as shown in Fig. S1B-F.
We extracted a mean value of $R_{\text{2t}}$ from each plot shown
in Sup 1F, averaged over the area in the $\left(B,V_{\text{bg}}\right)$
phase-space corresponding to each state. $R_{\text{2t}}$ of the unpolarized
state approaches the quantized value at long $L$ and goes down as
$L$ decrease, we fitted the data set using an exponential fit: $R(x)=\left(R\left(0\right)-R\left(\infty\right)\right)e^{-x/l_{0}}+R\left(\infty\right)$,
where $R\left(0\right)=20\pm13\,\text{k}\Omega$ is the value at zero
distance, $R\left(\infty\right)=38.2\pm0.3\,\text{k}\Omega$ is the
value at infinite distance and $l_{0}=2.1\pm0.8\,\mu\text{m}$ is
the typical equilibration length. Note that the zero distance resistance
is in agreement with the model presented in \citep{MacDonald1990}
for noninteracting edge channels where $R_{\text{2t}}=\left(4/3\,e^{2}/h\right)^{-1}=19.4\,\mathrm{k}\Omega$

\textcolor{black}{}
\begin{figure}[H]
\centering{}\textcolor{black}{\includegraphics[width=0.8\textwidth]{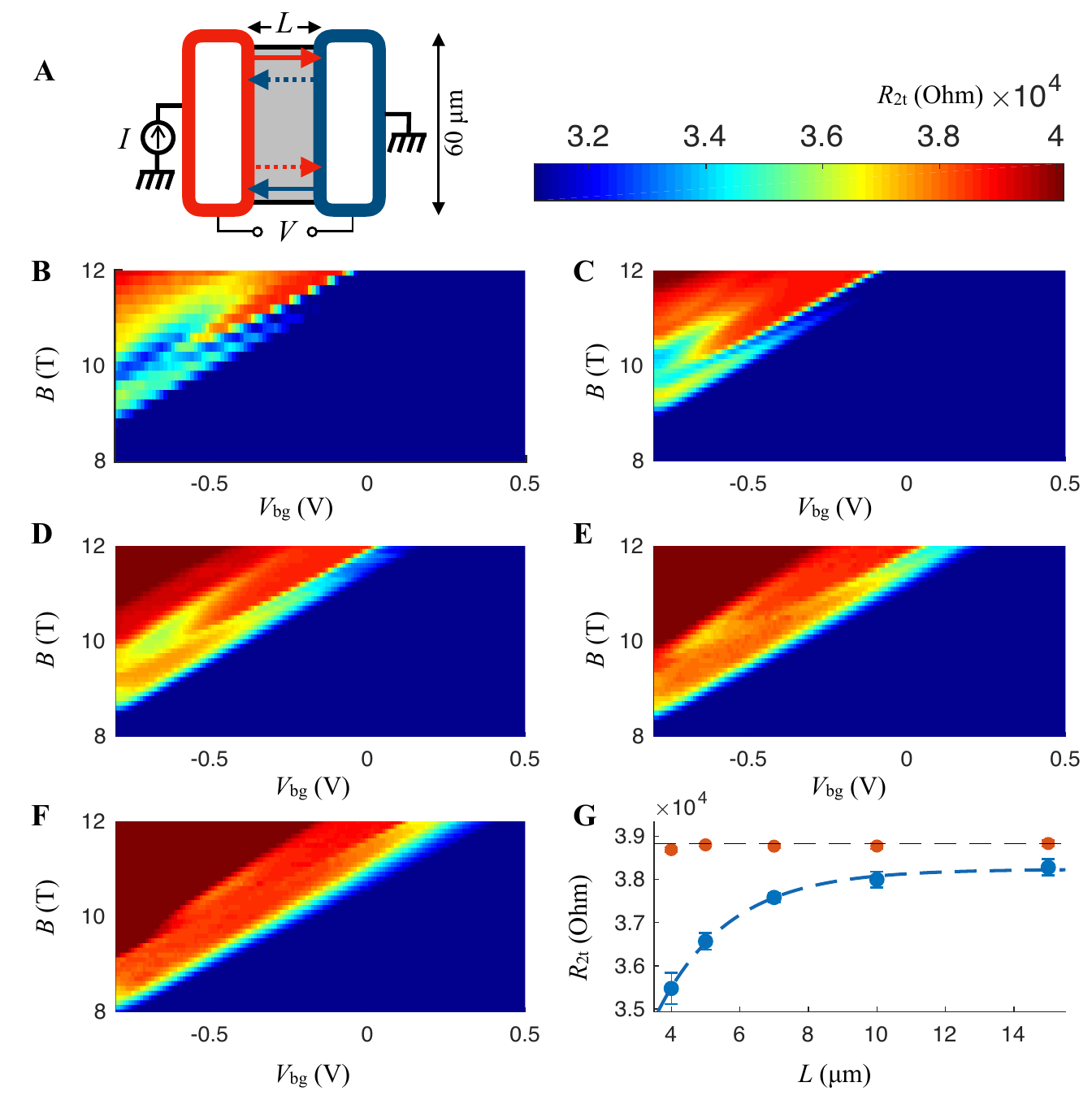}\caption{\textbf{Two-terminal magneto-resistance versus backgate voltage for
60 $\mu$m wide sample and 4 - 15 $\mu$m long at 40mK. (A) }Sketch
of two-terminal scheme. 1 nA was sourced (red contact) and drained
to ground (blue contact). Downstream chiral edge states illustrated
by solid arrows and upstream edge channels illustrated by dashed arrows.\textbf{
(B) }4$\mu$m, \textbf{(C) }5$\mu$m, \textbf{(D) }7$\mu$m, \textbf{(E)
}10$\mu$m, \textbf{(F) }15$\mu$m. \textbf{(G) }mean resistance of
$\nu=2/3$ unpolarized state (blue circles) and polarized state (orange
circles, error bars are smaller than the circle size for some points),
dashed black line represents the quantized value $\left(2/3\,e^{2}/h\right)^{-1}$
and the blue dashed line is an exponential fit $R(x)=\left(R\left(0\right)-R\left(\infty\right)\right)e^{-x/l_{0}}+R\left(\infty\right)$,
where $R\left(0\right)=20\pm13\,\text{k}\Omega$ , $R\left(\infty\right)=38.2\pm0.3\,\text{k}\Omega$
and $l_{0}=2.1\pm0.8\,\mu\text{m}$}
}
\end{figure}

\subsubsection{\textcolor{black}{Corbino geometry measurement}}

\textcolor{black}{To probe directly the conductance of the bulk we
designed a sample having a Corbino geometry as shown in Fig. S2A,
the outer $r_{out}$ and inner $r_{in}$ radius are respectively 14
and 6 microns. Voltage (20 $\mu$V) is sourced in the inner contact
and current is measured at the outer contact. The two terminal conductivity
$\sigma=1/2\pi\times\log(r_{out}/r_{in})\times(I/V)$ is plotted in
Fig. S2B. We can clearly see that both polarized and unpolarized $\nu=2/3$
states are in a non-dissipative state. }

\textcolor{black}{}
\begin{figure}[H]
\centering{}\textcolor{black}{\includegraphics[height=0.3\textwidth]{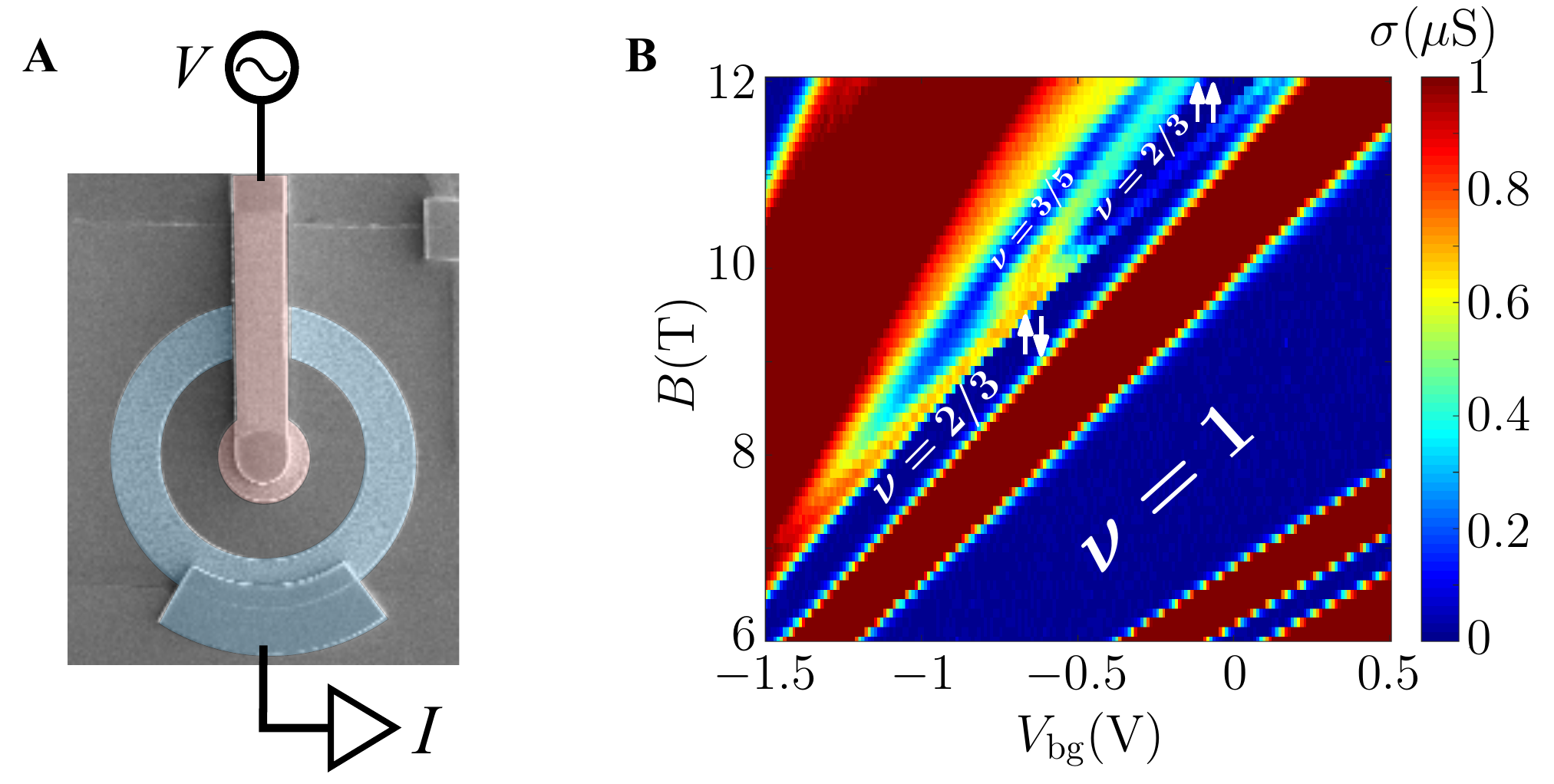}\caption{\textbf{Corbino geometry sample}. \textbf{(A) }Fake color SEM image
of the corbino sample together with the measurement scheme, the blue
colored ring shape ohmic contact has an inner radius of 14 $\mu\text{m}$
and is connected to cold ground. The red colored part include a disc
shape ohmic contact with radius of 6 $\mu\text{m}$ and is connect
to an air bridge. \textbf{(B) }Magneto-conductance measurement versus
the backgate voltage. It is clear that both $\nu=2/3$ polarized and
unpolarized are in a non dissipative state showing that the bulk is
incompressible in both cases.}
}
\end{figure}

\textcolor{black}{\newpage}

\subsubsection{\textcolor{black}{Landauer-Buttiker formalism }}

\textcolor{black}{Considering the system presented in Fig. S3A and
Fig. S3B, one can calculate the upstream and downstream conductance
using Landauer-Buttiker formalism \citep{Buttiker1986,Datta1995}: }

\textcolor{black}{
\[
\left(\begin{array}{c}
V_{1}\\
V_{2}\\
V_{3}
\end{array}\right)=[R]\left(\begin{array}{c}
I_{1}\\
I_{2}\\
I_{3}
\end{array}\right)=\left(\begin{array}{ccc}
R_{11} & R_{12} & R_{13}\\
R_{21} & R_{22} & R_{23}\\
R_{31} & R_{32} & R_{33}
\end{array}\right)\left(\begin{array}{c}
I_{1}\\
I_{2}\\
I_{3}
\end{array}\right)
\]
where}

\textcolor{black}{
\[
[R]=\left(\begin{array}{ccc}
G_{12}+G_{13}+G_{14} & -G_{12} & -G_{13}\\
-G_{21} & G_{21}+G_{23}+G_{24} & -G_{23}\\
-G_{31} & -G_{32} & G_{31}+G_{32}+G_{34}
\end{array}\right)^{-1}
\]
Inversing the matrix and replacing $G_{41}=G_{12}=G_{23}=G_{34}\equiv G_{d}$,
$G_{14}=G_{21}=G_{32}\equiv G_{u}$ and $G_{13}=G_{31}=G_{24}=0$
one finds the downstream resistance, which is the voltage measured
downstream divided by the current:}

\textcolor{black}{
\[
R_{12}=\frac{V_{1}}{I_{2}}\equiv R_{d}=\dfrac{G_{d}}{G_{u}^{2}+G_{d}^{2}}
\]
and the upstream resistance which is the voltage measured upstream
divided by the current:}

\textcolor{black}{
\[
R_{21}=\frac{V_{2}}{I_{1}}\equiv R_{u}=\dfrac{G_{u}}{G_{u}^{2}+G_{d}^{2}}
\]
one can find the upstream and downstream conductances:}

\textcolor{black}{
\begin{align*}
G_{d}= & \frac{R_{d}}{R_{d}^{2}+R_{u}^{2}}
\end{align*}
}

\textcolor{black}{
\begin{align*}
G_{u}= & \frac{R_{u}}{R_{d}^{2}+R_{u}^{2}}
\end{align*}
}

\textcolor{black}{In Fig. S3C and D, we plotted $R_{u}$ and $R_{d}$
directly from the 3-terminal measurement schemes presented in Fig.
S3A and B respectively. We found $R_{u}=1.42\ \text{k}\Omega\pm0.07\ \text{k}\Omega$
and $R_{d}=37.6\ \text{k}\Omega\pm0.1\ \text{k}\Omega$ where value
and the errors are the mean and standard deviation of the 2/3 unpolarized
region. Next, we plot the upstream and downstream conductances in
Fig. S3E and F, where we get the a mean value for the conductances
in the unpolarized region $G_{u}=0.026\,e^{2}/h\pm0.001\,e^{2}/h$
and $G_{d}=0.687\,e^{2}/h\pm0.002\,e^{2}/h$.}

\textcolor{black}{Since not all of the contacts have the same distance
from one another and since we claim the conductance of the 2/3 unpolarized
state depend on the distance, we also made another calculation. Taking
into consideration different conductance between contacts 3 and 4
which has much larger distance $\sim70$$\mu\text{m}$ where the edge
modes are fully equilibrated. Solving similar equations as above,
plugging the same values of $R_{u}$ and $R_{d}$ and taking the equilibrated
quantized value for $G_{34}=2/3\,e^{2}/h$ we get the same values
for $G_{u}$ and $G_{d}$ as the approximated ones up to our measurement
error.}

\textcolor{black}{}
\begin{figure}[H]
\centering{}\textcolor{black}{\includegraphics[height=0.8\textwidth]{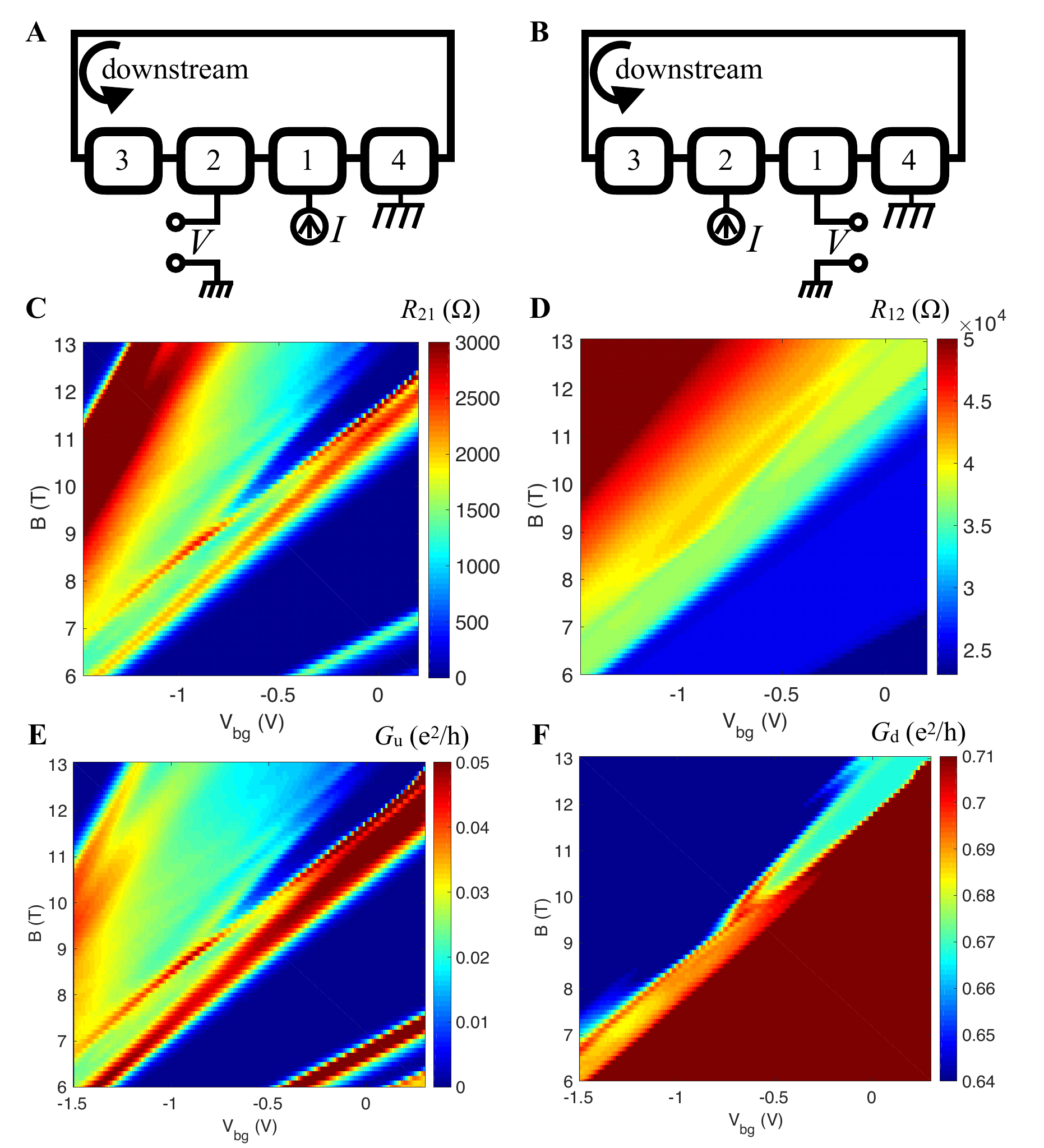}\caption{\textbf{Landauer-Buttiker analysis. (A) }3-terminal upstream measurement
scheme, 1 nA is sourced at contact 1 while voltage was measured at
contact 2, upstream. Contact 4 was grounded and contact 3 was floating
and wasn't in use. \textbf{(B) }3-terminal downstream measurement
scheme, 1 nA is sourced at contact 2 while voltage was measured at
contact 1, downstream. Contact 4 was grounded and contact 3 was floating
and wasn't in use. \textbf{(C)} upstream resistance measurement for
the scheme presented in (A).\textbf{ (D)} downstream resistance measurement
for the scheme presented in (B). \textbf{(E) }upstream conductance
calculated using L-B formalism $G_{u}=\frac{R_{u}}{R_{d}^{2}+R_{u}^{2}}$.
\textbf{F }downstream conductance calculated using L-B formalism $G_{d}=\frac{R_{d}}{R_{d}^{2}+R_{u}^{2}}$.}
}
\end{figure}

\textcolor{black}{\newpage}

\subsubsection{\textcolor{black}{Current conservation }}

\textcolor{black}{In the two QPCs configuration shown in Fig. S4C,
we sourced 1 nA in contact $S$ and measured current in all of the
drains separately. In Fig. S4A, We plot the current measured in the
downstream drains $D1+D2$ as function of the two QPCs split gate
voltages. In Fig. S4B, we plot the current measured in drain $D3$.
We observe a reduction of the current in the downstream drains when
upstream current is observed in $D3$. Overall yielding total current
conservation in all drains $D1$, $D2$ and $D3$.}

\textcolor{black}{}
\begin{figure}[H]
\centering{}\textcolor{black}{\includegraphics[height=0.6\textwidth]{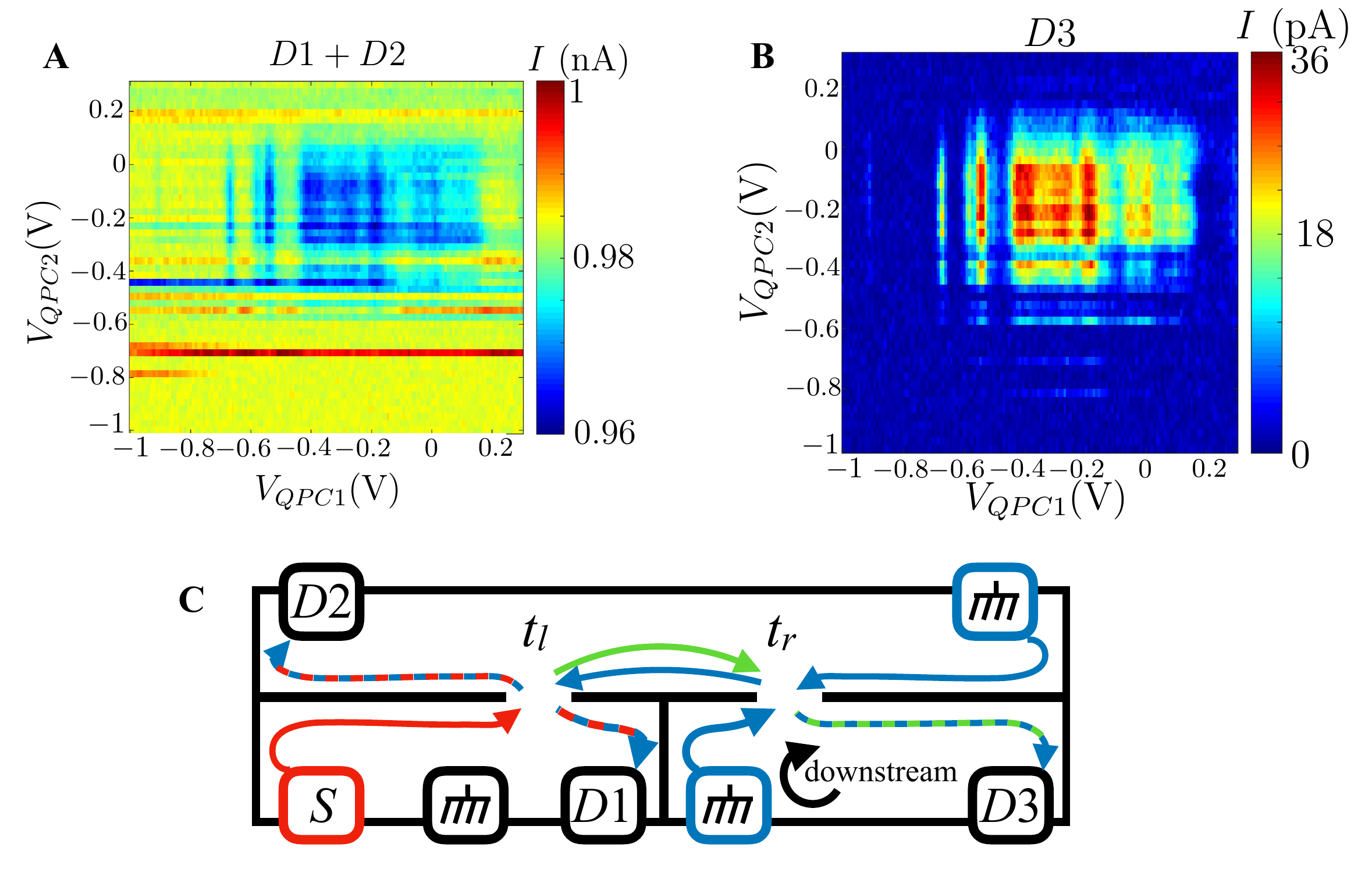}\caption{\textbf{Current conservation in two-point-contacts configuration}.\textbf{
(A)} Sum of the current measured in $D1+D2$ as function of the split
gate voltages.\textbf{ (B)} current measured in $D3$ as function
of the split gate voltages. \textbf{(C) }Two successive quantum point
contacts setup. Current is sourced in $S$ (red), flowing downstream
to the left QPC, there it is split to downstream (red/blue) and upstream
(green) charge currents. The unequilibrated upstream current reaches
the second QPC and turns back to downstream (green/blue) - measured
in $D3$. Distance between the two QPC is 700nm and the distance between
the QPC and the nearest ohmic contact is 30 $\mu$m (sketch not to
scale).}
}
\end{figure}

\subsubsection{Toy model for upstream current in two QPCs setup}

This toy model predicts the amount of current flowing into $D3$ in
the presence of upstream current. We modeled the two QPCs at half
transmission by two metallic ohmic contact, which have equilibrium
voltages $V_{l}$ and $V_{r}$ as shown in Fig. S5. Current $I$ sourced
in $S$ (red) reaches the left contact (green). From the left contact
we get one equilibrated downstream mode flowing long distance to $D1$
with conductance $2/3\ e^{2}/h$, another equilibrated downstream
mode flowing long distance to $D2$ with conductance $2/3\ e^{2}/h$
and one unequilibrated upstream mode with conductance $G_{u}$ flowing
a short distance (700 nm) to the right contact. Current conservation
on both contact yields:

\begin{align*}
I+V_{r}G_{d}= & V_{l}\left(2G_{2/3}+G_{u}\right)\\
V_{l}G_{u}= & V_{r}\left(G_{d}+G_{2/3}\right)
\end{align*}
where $G_{d}$ and $G_{u}$ are the unequilibrated downstream and
upstream conductance and $G_{2/3}=2/3\ e^{2}/h$ is the equilibrated
conductance. The current measured in $D3$ is $I_{3}=V_{r}G_{2/3}$ 

\begin{align*}
I_{3}= & \frac{G_{u}I}{2G_{2/3}+2G_{d}+G_{u}}
\end{align*}

For example, for Macdonald's model \citep{MacDonald1990} $G_{d}=e^{2}/h$,
$G_{u}=1/3e^{2}/h$ and $G_{2/3}=2/3e^{2}/h$ we get $I_{3}/I\approx0.09$,
about three times higher value than what we measured. We didn't take
into consideration any specific mechanism which convert dissipation
at the QPC to upstream modes, furthermore, one can take into account
different transmissions in the QPCs, where the model presented here
takes only half transmission. The highest signal of upstream current
we measured in correspondence with the highest dissipation occurring
in a QPCs, where $t_{l}=t_{r}=1/2$.

\begin{figure}[H]
\centering{}\includegraphics[width=0.6\textwidth]{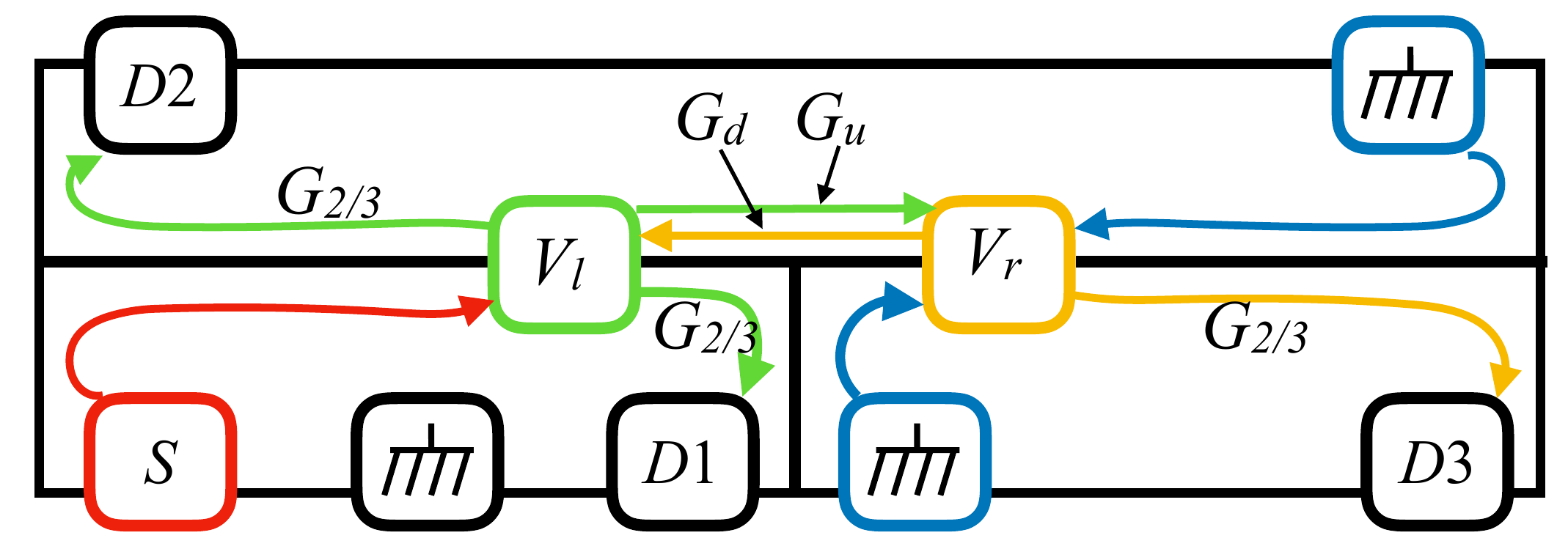}\caption{\textbf{Toy model sketch. }The two QPCs were replaced by ohmic contacts
which acquire a voltage $V_{l}$ and $V_{r}$. The short distance
between the two ohmic contacts (700 nm) give rise to equilibrated
edge modes propagating upstream and downstream flowing between the
two. Other edge modes are in full equilibration due to long distance,
above 30 $\mu$m (sketch not to scale).}
\end{figure}

\bibliographystyle{plain}

\textcolor{black}{}